\documentclass[prb,preprintnumbers,superscriptaddress,showpacs,twocolumn,balancelastpage,floatfix]{revtex4}

\usepackage{amsmath,amssymb,amsfonts}
\usepackage{bm,graphicx,color}

\newcommand{\TR}{\text{Tr}}

\newcommand{\Vk}{{\bm k}}
\newcommand{\Ks}{{{\bm k}\sigma}}
\newcommand{\tb}{{\bar t}}

\newcommand{\mydagger}{{\dagger}}
\newcommand{\phdagger}{{\phantom{\dagger}\!}}

\newcommand{\ket}[1]{|{#1}\rangle}

\newcommand{\expval}[1]{\langle{#1}\rangle}
\newcommand{\CS}{\mathcal{S}}

\newcommand{\delC}{\delta_\CC}
\newcommand{\intC}{\int_\CC}
\newcommand{\tmin}{t_{\text{min}}}
\newcommand{\tmax}{t_{\text{max}}}
\newcommand{\CC}{\mathcal{C}}
\newcommand{\TC}{\text{T}_{\CC}}

\newcommand{\convz}{\ast}

\newcommand{\ret}{{\text{R}}}
\newcommand{\adv}{{\text{A}}}
\newcommand{\mat}{{\text{\tiny M}}}
\newcommand{\tv}{{\makebox{$\neg$}}}
\newcommand{\vt}{{\reflectbox{$\neg$}}}
\newcommand{\les}{<}

\newcommand{\Ucdyn}{{U_{\text{dyn}}}}
\newcommand{\sss}{_{\alpha\alpha'}}
\newcommand{\YY}{Y}
\newcommand{\KK}{K}

\begin{document}

  \title{Interaction quench in the Hubbard model: Relaxation of the spectral function and the optical conductivity}

  \author{Martin Eckstein}
  \affiliation{Theoretical Physics, ETH Zurich, 8093 Zurich, Switzerland}

  \author{Marcus Kollar}
  \affiliation{Theoretical Physics III, Center for Electronic
    Correlations and Magnetism, Institute for Physics, University of
    Augsburg, 86135 Augsburg, Germany}
  
  \author{Philipp Werner}
  \affiliation{Theoretical Physics, ETH Zurich, 8093 Zurich, Switzerland}  
  
  \date{\today}

  \begin{abstract}
    We use non-equilibrium dynamical mean-field theory in combination
    with a recently developed Quantum Monte Carlo impurity solver to study the
    real-time dynamics of a Hubbard model which is driven out of
    equilibrium by a sudden increase in the on-site repulsion $U$. We
    discuss the implementation of the self-consistency procedure and
    some important technical improvements of the QMC method. The exact
    numerical solution is compared to iterated perturbation theory,
    which is found to produce accurate results only for weak
    interaction or short times. Furthermore we calculate the spectral
    functions and the optical conductivity from a Fourier transform on 
    the finite Keldysh contour, for which the numerically accessible 
    timescales allow to resolve the formation of Hubbard bands and a 
    gap in the strongly interacting regime. The spectral function, 
    and all one-particle quantities that can be calculated 
    from it, thermalize rapidly at the transition between  
    qualitatively different weak- and strong-coupling relaxation regimes.
  \end{abstract}


  \maketitle
  
  \noindent
  \section{Introduction} 

  The recent realization of a Mott insulating state of
  repulsively interacting fermions in trapped ultracold
  atoms\cite{Joerdens2008a,Schneider2008a} opens the door to 
  controlled studies of the non-equilibrium properties of fermionic
  lattice models. 
  At the same time, the relaxation dynamics of strongly correlated electron systems is
  starting to be explored experimentally through femtosecond
  spectroscopy.\cite{Iwai2003a,Perfetti,Kawakami2009,Wall2009} 
  Dynamical mean-field theory\cite{Georges1996} (DMFT)
  is a promising tool to approach these challenging issues from the
  theoretical side. The DMFT formalism is based on 
  the mapping of  a lattice model to a quantum impurity model.
  This approximation is based on a purely spatial
  argument  which becomes exact in the limit of  infinite dimensions.\cite{Metzner1989a} 
  On the one hand this fact
  makes DMFT a nonperturbative method which can capture, e.g.,
  the local Mott physics of the Hubbard model. On the other hand, it
  implies that DMFT can be formulated equally well in imaginary and
  real time, and hence the method can be applied to both equilibrium
  and nonequilibrium situations.\cite{Schmidt2002}

  A number of authors have employed the non-equilibrium DMFT framework
  to study dynamical properties of the Falicov-Kimball model, which is
  a variant of the Hubbard model in which only one spin species can
  hop between lattice sites.  Despite this simplification, the
  Falicov-Kimball model exhibits a relatively rich phase diagram with
  metallic, Mott-insulating, and charge-ordered
  phases.\cite{Freericks2003a} Its most attractive feature in the
  present context is that the associated quantum impurity model in
  DMFT can be solved analytically or numerically by simple
  means,\cite{Brandt89} which provides reliable access to the
  long-time dynamics.  Both the transient dynamics after the sudden
  switching-on of a static electric
  field\cite{Freericks2006a,Freericks2008b,Tran2008a} and, using a
  combined Floquet and DMFT formalism, the non-equilibrium steady
  state in the presence of an alternating or constant
  field\cite{Tsuji2008a,Joura2008a,Tsuji2009a} were calculated.  The
  evolution of the momentum distribution and double occupation after
  an interaction quench, i.e., a sudden change in the interaction
  parameter, was studied in Ref.~\onlinecite{Eckstein2008a}, where it
  was also shown that for the Falicov-Kimball model these quantities
  do not thermalize. This is a consequence of the immobility of one
  spin species and the resulting quadratic form of the Hamiltonian for
  the mobile other spin species.

  A more realistic model for the description of correlated electron
  systems and interacting fermions in optical lattices is the Hubbard
  model,
  \begin{align}
    \label{hubbard}
    H(t)
    &=
    \sum_{ij\sigma}
    V_{ij}
    c_{i\sigma}^{\mydagger}
    c_{j\sigma}^{\phdagger}
    +
    U(t)
    \sum_{i}
    \big(n_{i\uparrow}\!-\!\tfrac12\big)
    \big(n_{i\downarrow}\!-\!\tfrac12\big)
    ,
  \end{align}
  which describes fermions of spin one half which hop on a lattice
  with hopping amplitude $V_{ij}$ and interact on each site with a
  repulsion energy $U$. 
  An interaction quench has so far been experimentally realized in 
  the bosonic version of the Hubbard model.\cite{Greiner2002b} 
  To describe the corresponding situation in the fermionic model,
  we allow for a time-dependent interaction $U(t)$ in Eq.~(\ref{hubbard}).

  Even after the mapping to a single-site model, 
  the solution of the Hubbard model within nonequilibrium DMFT
  requires the calculation of the time evolution of an interacting
  many-body system.  In a previous publication\cite{Eckstein2009a} we
  employed a recently developed diagrammatic impurity
  solver\cite{Werner09} to compute the time evolution after an
  interaction quench over a wide parameter regime within DMFT.  The
  numerical simulations confirmed an analytical flow equation analysis
  for quenches to small $U$,\cite{Moeckel2008} which showed that in
  the limit $U\rightarrow 0$ the system is trapped in a nonthermal
  metastable intermediate state, a phenomenon known as
  prethermalization.\cite{Berges2004a} We identified a similar
  trapping phenomenon for quenches to very large interactions. Most
  interestingly, these two prethermalization regimes are separated by
  a well-defined ``critical'' interaction $\Ucdyn$, where instead of a
  trapping in either of the nonthermal states a fast thermalization
  is observed. In Ref.~\onlinecite{Eckstein2009a} these qualitatively different
  regimes were demonstrated on the basis of an analysis of the 
  momentum distribution and the double occupancy.
  
  Relaxation to
  a thermal state is often difficult to establish numerically because
  the time evolution must be studied on long timescales.  In general
  thermalization is expected for interacting systems (in the weak
  sense that the expectation value of a large class of observables
  approaches the thermal expectation value in the long-time limit).
  In exactly solvable systems, however,  thermalization is often prevented by
  integrability,\cite{Rigol2007a,Cazalilla06,Eckstein2008a,Kollar2008}
  while its mechanism for nonintegrable systems is currently under
  debate.\cite{Manmana07,Rigol2008a,Roux2009a,Biroli2009a,Rigol2009b,Santos2009}
  The existence of two separate relaxation regimes is similar to what
  was found for Heisenberg spin chains\cite{Barmettler2009a} and the
  one-dimensional Bose-Hubbard model.\cite{Kollath2007} 

  The purpose of the present work is twofold: First, we want to
  explain in some detail the machinery behind our Quantum Monte Carlo
  (QMC) calculation of the Hubbard model in nonequilibrium DMFT.  We
  will briefly review the DMFT formalism (Sec.~\ref{sec-noneqdmft})
  and the diagrammatic Monte Carlo method (Sec.~\ref{sec-qmc}),
  discuss some important tricks which improve the efficiency of the
  Monte Carlo sampling, and then present in detail the solution of the
  DMFT self-consistency 
  equations based on the exact equation of motion approach
  (Sec.~\ref{sec-qmcdmft}).  The QMC solution of DMFT is finally used
  to discuss the validity of the nonequilibrium generalization of the
  iterated perturbation theory (Sec.~\ref{dmft-sec-ipt}). The second
  purpose of this paper is to further analyze the main finding of
  Ref.~\onlinecite{Eckstein2009a}, namely a fast thermalization after
  a quench from $U=0$ to $U_\text{dyn}$, with additional data for the
  momentum distribution, the spectral function, and the optical
  conductivity (Sec.~\ref{sec-results}). In particular we find that at
  $U_\text{dyn}$ the retarded non-equilibrium Green function relaxes
  to the appropriate equilibrium Green function within the numerical
  accuracy, establishing thermalization of all one-particle quantities
  that can be calculated from it.

  \section{Nonequilibrium DMFT}
  \label{sec-noneqdmft}
  \subsection{Contour-ordered Green functions}
  \renewcommand{\tmin}{0}

\begin{figure}
    \begin{center}
      \includegraphics[angle=0,width=0.9\columnwidth]{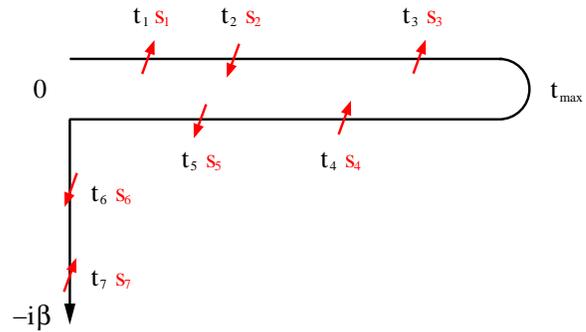}
      \caption{
        The L-shaped contour $\CC$ for the description of transient 
        nonequilibrium states with initial state density matrix 
        $\propto e^{-\beta H(0)}$. Arrows show a possible Monte Carlo 
        configuration corresponding to perturbation
        order $n=7$ and $n_+=3$, $n_-=2$, $n_\beta=2$ (cf.~Sec.~\ref{sec-qmc}).
      }
      \label{contour}
    \end{center}
  \end{figure}

  In the following section we set up the framework for the
  investigation of a rather general class of nonequilibrium
  situations. We assume that the system of interest is initially prepared in
  thermal equilibrium. For times $t>\tmin$ it is then acted on by some perturbation, 
  but there is no coupling to external heat or particle
  reservoirs. Technically, this setup implies that the time evolution
  is unitary and captured by a time-dependent Hamiltonian, but all
  results must be averaged over initial states according to the
  grand-canonical density matrix $\rho_0$ $=$ $e^{-\beta
    H(0)}/\TR[e^{-\beta H(0)}]$ at temperature $T=1/\beta$. The
  conventional approach to this kind of nonequilibrium situation
  within many-body theory is based on the use of contour-ordered 
  Keldysh Green functions,\cite{keldyshintro}
  \begin{equation}
    \label{ky-cntr}
    G_{\alpha\alpha'}(t,t') = 
    -i 
    \expval{\TC\, \hat c_\alpha(t) \hat c_{\alpha'}^\mydagger(t')}
    \,,
  \end{equation}
  where the time arguments $t$ and $t'$ lie on the L-shaped contour
  $\CC$ that runs from $\tmin$ to some time $\tmax$ (i.e., the largest
  time of interest) on the real time axis, back to $\tmin$, and
  finally to $-i\beta$ along the imaginary time axis
  (Fig.~\ref{contour}).  Here and in the following, operators with hat
  are in Heisenberg notation with respect to the time-dependent
  Hamiltonian [on the imaginary branch, $H=H(\tmin)$, and $\hat
  c(-i\tau) = e^{\tau H(\tmin)} c e^{-\tau H(\tmin)}$], and
  $\expval{\cdot}$ $=$ $\TR[\rho_0 \cdot]$ is the expectation value
  taken in the initial equilibrium state. The contour-ordering $\TC$
  exchanges the order of two operators $A(t_1)$ and $B(t_2)$ in a
  product $A(t_1)B(t_2)$ if and only if $t_2$ appears later on the
  contour than $t_1$, with an additional minus sign if the exchange
  involves an odd number of Fermi operators. The order of time
  arguments along $\CC$ is indicated by the arrow in
  Fig.~\ref{contour}, which points from ``earlier'' to ``later''
  times. Contour-ordered Green functions were first introduced by
  Keldysh\cite{Keldysh1964a} in order to generalize Wick's theorem and
  diagrammatic perturbation theory to nonequilibrium physics. The
  extension of the original Keldysh formalism to the L-shaped contour
  $\CC$, which has numerous applications in nonequilibrium many-body
  theory,\cite{Bonitz2003a} becomes important whenever correlations
  between the initial state at $t=0$ and time $t>0$ cannot be
  neglected.\cite{Danielewicz1984a,Wagner1991a}

  The contour-ordered Green function (\ref{ky-cntr}) is related to a
  number of real and imaginary-time Green functions, which we list in
  the following paragraph for later reference. When both time
  arguments are on the imaginary branch, Eq.~(\ref{ky-cntr}) reduces
  to the Matsubara Green function of the initial equilibrium state,
  \begin{equation}
    \label{ky-cntr-mat}
    G_{\alpha\alpha'}^\mat(\tau,\tau') =
    G_{\alpha\alpha'}(-i\tau,-i\tau').
  \end{equation}
  Because the Hamiltonian is constant on the vertical branch of $\CC$
  and commutes with the initial state density matrix,
  $G_{\alpha\alpha'}^\mat(\tau,\tau')$ is translationally invariant in
  imaginary time, such that we can introduce the usual Matsubara frequency
  representation,
  \begin{subequations}
    \label{ky-mat}
    \begin{align}
      \label{ky-mat-tt}
      G^\mat(\tau,\tau')
      =
      \frac{i}{\beta}
      \sum_n e^{i\omega_n(\tau'-\tau)}
      g^\mat(i\omega_n),
      \\
      \label{ky-mat-w}
      g^\mat(i\omega_n)
      =
      -i\int\limits_0^\beta \!d\tau\, e^{i\omega_n\tau} \,G^\mat(\tau,0).
    \end{align}
  \end{subequations}
  On the other hand, when both time arguments are real, one obtains
  the lesser, retarded, and advanced Green functions,
  \begin{eqnarray}
    \label{ky-cntr-les}
    G_{\alpha\alpha'}^\les(t,t') 
    &\equiv& G_{\alpha\alpha'}(t_+,t'_-)
    = 
    i\expval{\hat c_{\alpha'}^\mydagger(t')\hat c_\alpha(t) }_0\\
    \label{ky-cntr-ret}
    G_{\alpha\alpha'}^\ret(t,t') 
    &\equiv& 
    \Theta(t-t') [G_{\alpha\alpha'}(t_-,t'_+) -G_{\alpha\alpha'}(t_+,t'_-)]\nonumber\\
    \label{ky-cntr-ret1}
    &=& 
    -i\Theta(t-t')\expval{\{\hat c_{\alpha'}^\mydagger(t'),\hat c_\alpha(t)\} },
    \\
    \label{ky-cntr-adv}
    G_{\alpha\alpha'}^\adv(t,t') 
    &\equiv& 
    \Theta(t'-t) [G_{\alpha\alpha'}(t_+,t'_-) -G_{\alpha\alpha'}(t_-,t'_+)]\nonumber\\
    &=& 
    i\Theta(t'-t)\expval{\{\hat c_{\alpha'}^\mydagger(t'),\hat c_\alpha(t)\} }.
  \end{eqnarray}
  The subscript of each real time-argument indicates whether it is on
  the upper ($+$) or lower ($-$) real-time branch of $\CC$. The lesser
  Green function is related to the occupation of states $\alpha$, to
  which its imaginary part reduces for $t=t'$ and $\alpha=\alpha'$. On
  the other hand, the retarded and advanced Green function are related
  to the spectral function, which will be discussed in more detail in
  Sec.~\ref{sec-results}. In addition to the real and imaginary time
  Green functions, the Green functions
  \begin{subequations}
    \label{ky-cntr-mix}
    \begin{align}
      \label{ky-cntr-tv}
      G_{\alpha\alpha'}^\tv(t,\tau) 
      &\equiv
      G_{\alpha\alpha'}(t_\pm,-i\tau),
      \\
      \label{ky-cntr-vt}
      G_{\alpha\alpha'}^\vt(\tau,t) 
      &\equiv
      G_{\alpha\alpha'}(-i\tau,t_\pm).
    \end{align}
  \end{subequations}
  with mixed time arguments encode the correlations between the
  initial state and times $t>0$.

  It follows from the cyclic property of the trace and the definition
  of the contour-ordering that the Green function (\ref{ky-cntr})
  satisfies an antiperiodic boundary condition on $\CC$ in both
  time-arguments,
  \begin{subequations}
    \label{ky-bc}
    \begin{align}
      G_{\alpha\alpha'}(0_+,t') &= - G_{\alpha\alpha'}(-i\beta,t'),\\
      G_{\alpha\alpha'}(t,0_+) &= - G_{\alpha\alpha'}(t,-i\beta).
    \end{align}
  \end{subequations}
  This boundary condition holds for all contour functions in this
  text, including those which have no simple definition in terms of
  Heisenberg operators.  Furthermore, the Green function
  (\ref{ky-cntr}) satisfies the hermitian symmetry
  \begin{subequations}
    \label{ky-cntr-sym}
    \begin{align}
      \label{ky-cntr-sym-adv}
      G^\ret_{\alpha\alpha'}(t,t') &= G^\adv_{\alpha'\alpha}(t',t)^*\\
      \label{ky-cntr-sym-les}
      G^\les_{\alpha\alpha'}(t,t') &= -G^\les_{\alpha'\alpha}(t',t)^*\\
      \label{ky-cntr-sym-mix}
      G^\tv_{\alpha\alpha'}(t,\tau) &= G^\vt_{\alpha'\alpha}(\beta-\tau,t)^*,
    \end{align}
  \end{subequations}
  which will be used frequently in the following.

  \subsection{Dynamical mean-field theory}

  In equilibrium DMFT  local correlation functions are
  obtained from a single-site impurity model subject to a
  self-consistency condition.\cite{Georges1996} The mapping of the
  lattice problem (\ref{hubbard}) onto the single-site problem is
  formally achieved by integrating out all lattice sites apart from
  one. A straightforward reformulation of this mapping for Green
  functions on the Keldysh contour\cite{Schmidt2002,Freericks2006a}
  makes DMFT applicable to nonequilibrium problems. The single-site
  action is then given by
  \begin{subequations}
    \label{dmft-action}
    \begin{align}
      \mathcal{S}
      &= \mathcal{S}_0 + 
      \int\limits_\CC\!dt\, h_{\text{loc}}(t),
      \\
      \mathcal{S}_0
      &=\!\!
      \sum_{\sigma=\uparrow,\downarrow}
      \int\limits_\CC \!\!dt\,dt'\,c_\sigma^\mydagger(t)
      \Lambda_\sigma(t,t')c_\sigma(t'),
    \end{align}
  \end{subequations}
  where $\intC dt = \int_0^{\tmax} \!dt_+ \!-\! \int_0^{\tmax} \!dt_-
  \!-i \int_0^{\beta} d\tau$ is the integral along $\CC$,
  \begin{equation}
    \label{dmft-hloc}
    h_{\text{loc}}(t)
    =
    U(t)\big(n_{\uparrow}-\tfrac12\big)\big(n_{\downarrow}-\tfrac12 \big)
  \end{equation}
  is the local interaction of the Hamiltonian, and $\mathcal{S}_0$
  describes the hybridization of the site with an environment that is
  determined self-consistently by the DMFT procedure. In the following
  we consider only homogeneous paramagnetic phases, such that
  $\Lambda_\sigma$ does not depend on the lattice site or spin $\sigma$.

  The local Green function for action (\ref{dmft-action}) is given by
  \begin{equation}
    \label{dmft-gloc}
    G_{\sigma} (t,t')
    =
    -i \expval{c_\sigma(t) c_\sigma^\dagger(t')}_{\CS},
  \end{equation}
  where operators without a hat are in the interaction picture with
  respect to $\mu (n_\uparrow+n_\downarrow)$, and the notation
  \begin{equation}
    \expval{\,\cdots\,}_\CS
    =
    \frac{\TR [e^{-\beta\mu(n_\uparrow+n_\downarrow)}\TC \exp(-i\CS)\,\cdots\,]}
    {\TR [e^{-\beta\mu(n_\uparrow+n_\downarrow)}\TC\exp(-i\CS)]}
  \end{equation}
  is used. In general, the computation of $G_{\sigma} (t,t')$ is a
  complicated nonequilibrium many-body problem. For this reason,
  nonequilibrium DMFT has so far been applied mostly to the
  Falicov-Kimball model, where the single-site problem can be reduced
  to a quadratic one and thus becomes exactly solvable either
  numerically\cite{Freericks2006a,Tran2008a} or
  analytically.\cite{Eckstein2008a} In the present paper, just as in
  Ref.~\onlinecite{Eckstein2009a}, we investigate the Hubbard model
  and solve the single-site problem using the weak-coupling continuous
  time Monte Carlo algorithm,\cite{Werner09} which will be described
  below (Sec.~\ref{sec-qmc}).

  The local self-energy is then defined by the Dyson equation
  \begin{align}
    \label{dmft-dysonloc}
    [(G_{0,\sigma}^{-1} - \Sigma_{\sigma}) \convz G_{\sigma}](t,t') 
    = \delC(t,t'),
  \end{align}
  where the noninteracting ($U=0$) single-site Green function 
  and its inverse are  given by
  \begin{subequations}
    \label{dmft-g0loc}
    \begin{align}
      \label{dmft-g0loc-def}
      G_{0,\sigma} (t,t')
      &=
      -i \expval{c_\sigma(t) c_\sigma^\dagger(t')}_{\CS_0},
      \\
      G_{0,\sigma}^{-1}(t,t') 
      &= \delC(t,t')(i\partial_t + \mu) - \Lambda_{\sigma}(t,t'),\label{dmft-g0loc-inv}
    \end{align}
  \end{subequations}
  respectively. Here we introduced the notation $[a \convz b](t,t')$
  $=$ $\intC d\tb \,a(t,\tb)\,b(\tb,t')$ for the convolution of two
  contour functions, and the contour-delta function $\delC(t,t')$ is
  defined such that
  \begin{equation}
    \intC \! d\tb \, f(\tb)\delC(\tb,t) = f(t)
  \end{equation}
  for any contour function $f$, i.e., 
  $\delC(t,t')=\pm\delta(t-t')$ if  $t$ and $t'$ both on the
  upper or lower real branch of $\CC$, and 
  $\delC(-i\tau,-i\tau')=i\delta(\tau-\tau')$
  for time arguments on the vertical branch.
  Both the Dyson equation
  (\ref{dmft-dysonloc}) and the corresponding equation
  \begin{equation}
    \label{pseudodyson}
    [G_{0,\sigma}^{-1}\convz G_{0,\sigma}](t,t')  =  \delC(t,t'),
  \end{equation}
  for $G_{0,\sigma}$ are inhomogeneous integro-differential equations
  on the contour $\CC$ when Eq.~(\ref{dmft-g0loc-def}) is inserted.
  They have a unique solution because $G_{0,\sigma}$ and $G$ satisfy
  the boundary condition (\ref{ky-bc}). The solution of such integral
  equations on $\CC$ is discussed in detail in Sec.~\ref{sec-qmcdmft}.

  In order to determine the hybridization function $\Lambda_\sigma(t,t')$ one
  must equate the self-energy $\Sigma_\sigma(t,t')$ and the Green
  function $G_\sigma(t,t')$ of the single-site problem with the local
  self-energy $\Sigma_{jj\sigma}(t,t')$ and the local Green function
  $G_{jj\sigma}(t,t')$ of the lattice problem at the given site $j$,
  respectively,
  \begin{equation}
    \label{dmft-sce-general}
    G_{jj\sigma}(t,t')=G_{\sigma}(t,t'),\quad  \Sigma_{ij\sigma}(t,t')=\delta_{ij}\Sigma_{\sigma}(t,t').
  \end{equation}
  The latter two are related by the lattice Dyson equation,
  \begin{equation}
    \label{dmft-lattice-dyson-kspace}
    (i\partial_t+\mu-\epsilon_{\Vk})
    G_\Ks(t,t')
    -
    [\Sigma_\sigma \convz G_{\Ks}](t,t')
    =
    \delC(t,t'),
  \end{equation}
  which is stated here for the homogeneous case after Fourier
  transform with respect to lattice sites. In
  Eq.~(\ref{dmft-lattice-dyson-kspace}),
  \begin{equation}
    G_{\Vk\sigma}(t,t') = -i\expval{\TC \hat c_{\Vk\sigma}(t) \hat c_{\Vk\sigma}^\dagger(t')}
  \end{equation}
  is the momentum-resolved lattice Green function.  (For a Bravais
  lattice, $\Vk$ are quasimomenta and $\epsilon_\Vk$ are band
  energies, but more generally, $\expval{i|\Vk}$ and $\epsilon_{\Vk}$
  are eigenvectors and eigenvalues of the hopping matrix $V_{ij}$,
  respectively.)  The local Green function is given by the momentum
  sum
  \begin{equation}
    \label{localg}
    G_{\sigma}(t,t') = \sum_{\Vk} |\expval{j|\Ks}|^2 G_\Ks(t,t'),
  \end{equation}
  which closes the self-consistency.

  In the present paper we consider the case of a time-dependent
  interaction but no external fields. The hopping matrix elements are
  then independent of time, and the $\Vk$-summation in
  Eq.~(\ref{localg}) can be reduced to an integral over a single
  energy variable
  \begin{equation}%
    \label{lattice-dyson-gloc}
    G_\sigma(t,t')
    =
    \int\!d\epsilon\,\rho(\epsilon)G_{\epsilon\sigma}(t,t'),
  \end{equation}%
  involving the Green function $G_{\epsilon\sigma}(t,t')$ $=$
  $G_\Ks(t,t')|_{\epsilon_\Ks=\epsilon}$ and the local density of
  states $\rho(\epsilon)$ $=$ $\sum_{\Vk} |\expval{j|\Vk}|^2
  \delta(\epsilon-\epsilon_{\Vk})$ at an arbitrary site $j$. For the
  case of a semielliptic density of states,
  \begin{equation}
    \label{rho-semi}
    \rho(\epsilon) = \frac{\sqrt{4 V^2 -\epsilon ^2}}{2 \pi V},
  \end{equation}
  with quarter bandwidth $V$, which corresponds to nearest-neighbor
  hopping on the Bethe lattice\cite{Economou1979a,Georges1996} or a
  particular kind of long-range hopping on the hypercubic
  lattice,\cite{Bluemer2003a} one obtains a closed form expression for
  the Weiss field,\cite{Eckstein2008a}
  \begin{equation}
    \label{selfconsistency-bethe}
    \Lambda_\sigma(t,t') = V^2 G_\sigma(t,t').
  \end{equation}
  We will use this self-consistency equation for all results of this
  work, so that the solution of the DMFT equations is achieved by
  iteration of  Eqs.~(\ref{dmft-gloc}) and (\ref{selfconsistency-bethe}).

  \section{Real-time Monte Carlo method}
  \label{sec-qmc}

  The real-time evolution of the impurity model can be computed using
  the weak-coupling diagrammatic Monte Carlo method. More
  specifically, we employed the real-time version of the
  continuous-time auxiliary field algorithm (CTAUX)\cite{Gull08}
  which was discussed in detail in Ref.~\onlinecite{Werner09}. 
  In the following section we present the implementation of this algorithm on the L-shaped
  contour (Fig.~\ref{contour}) and then discuss some technical aspects
  which improve the efficiency of the method to the point where
  relevant timescales even in the strong-coupling regime become
  accessible.

  We start by expressing the partition function of the initial state as
  %
  \begin{eqnarray}
    Z &=& \text{Tr} \big[e^{-\beta\mu(n_\uparrow+n_\downarrow)}T_{\mathcal{C}}
    e^{-i\mathcal{S}}\big]\nonumber\\
    &=& e^{-\beta U/4-i\int_\mathcal{C} dt k(t)}\text{Tr} \big[e^{-\beta\mu(n_
      \uparrow+n_\downarrow)}T_{\mathcal{C}} e^{-i\mathcal{S}_k}\big]
    \label{identity}
  \end{eqnarray}
  %
  with $\mathcal{S}_k=\mathcal{S}-\int_\mathcal{C}dt (k(t)+U/4)$ and
  $k(t)\ne 0$. This (possibly time-dependent) shift in the action is
  introduced so that the interaction term can be decoupled using
  auxiliary Ising spin variables according to\cite{Rombouts99}
  %
  \begin{eqnarray}
    -h_\text{loc}(t)+k(t)+U/4
    &=& k(t)-U(n_{\uparrow} n_{\downarrow}-(n_{\uparrow}+n_{\downarrow})/
    2))\nonumber\\
    &=& k(t)/2\sum_{s=-1,1}e^{\gamma(t) s (n_{\uparrow}-n_{\downarrow})},
    \nonumber\\
    \cosh(\gamma(t))&=&1+U/(2k(t)).
    \label{decouple}
  \end{eqnarray}
  %
  Expansion of $e^{-i\mathcal{S}_k}$ in powers of
  $(h_\text{loc}(t)-k(t)- U/4)$ and subsequent auxiliary field
  decomposition leads to an expression of the partition function as a
  sum over all possible Ising spin configurations on the contour
  $\mathcal{C}$.  The weight of the Monte Carlo configurations $\{
  (t_{1}, s_1),(t_{2}, s_2), \ldots (t_{n}, s_n) \}$ (see illustration
  in Fig.~\ref{contour}) is obtained by evaluating the trace of the
  remaining noninteracting problem,
  \begin{widetext}
    \begin{eqnarray}
      w(\{ (t_1, s_1), \ldots (t_{n}, s_n) \})&=&
      (ik(t_1)dt/2)\ldots (ik(t_{n_+})dt/2)(-ik(t_{n_++1})dt/2)\ldots (-
      ik(t_{n_++n_-})dt/2)\nonumber\\
      &&\times (k(t_{n_++n_-+1})d\tau/2)\ldots (k(t_{n_++n_-+n_\beta})d\tau/2)
      \prod_\sigma \det N_\sigma^{-1},\label{weight}\\
      N_\sigma^{-1} &=& e^{\Gamma_\sigma}-iG_{0,\sigma}(e^{\Gamma_\sigma}-I).
    \end{eqnarray}
  \end{widetext}
  Here $n_\pm$ and $n_\beta$ denotes the number of Ising spins on
  the three branches of $\CC$, 
  $G_{0,\sigma}$ is the $(n_++n_-+n_\beta)\times(n_++n_-+n_\beta)$ matrix
  of bath Green functions (\ref{dmft-g0loc-def}) evaluated at the time arguments 
  defined by
  the Ising spins, and $e^{\Gamma_\sigma}=\text{diag}(e^{\gamma(t_1)
    s_1\sigma}, \ldots, e^{\gamma(t_n) s_{n} \sigma})$. A Monte Carlo
  sampling of all possible spin configurations can then be implemented
  based on the absolute value of these weights.

  The contribution of a specific configuration $c$ to the Green function is
  given by
  %
  \begin{eqnarray}
    G^c_\sigma(t,t')&=&G_{0,\sigma}(t,t')\nonumber\\
    &+&i\sum_{i,j=1}^n G_{0,\sigma}(t,t_{i})[(e^{\Gamma_\sigma}-1)N_
    \sigma]_{i,j}G_{0,\sigma}(t_{j}, t'),\nonumber\\
    \label{tildeG}
  \end{eqnarray}
  %
  so the Green function $G$ is obtained as the Monte Carlo average of $G^c$.

  The sign problem in this method grows exponentially with the average
  perturbation order on the real-time portion of $\CC$. To reach long times
  or strong interactions, it is therefore important to reduce this
  perturbation order as much as possible. In the particle-hole
  symmetric case, i.e., at half-filling and for a symmetric density of
  states, the parameter $k(t)$ of the algorithm can be chosen such
  that only even perturbation orders appear in the expansion. In fact,
  for
  \begin{equation}
    k(t)=-U/4
  \end{equation}
  we have $\gamma(t)=i\pi$, $e^{\gamma(t) s \sigma}=-1$ and hence the
  spin degree of freedom effectively disappears. The algorithm then 
  becomes the real-time version of Rubtsov's weak-coupling
  method\cite{Rubtsov05} for the particle-hole symmetric interaction
  term $h_ \text{loc}(t)=U(n_{\uparrow}-
  \frac{1}{2})(n_{\downarrow}-\frac{1}{2})$ with weight
  \begin{eqnarray}
    w_\text{even}(t_1, \ldots, t_n)&=&
    (-iUdt)^{n_+}(iUdt)^{n_-}(-Ud\tau)^{n_\beta}\nonumber\\
    &&\times \prod_\sigma \det \Big(iG_{0,\sigma}-\frac{1}{2}I\Big).
    \label{weight_Rubtsov}
  \end{eqnarray}
  (For a detailed discussion of the equivalence between the Rubtsov
  and CTAUX methods for the Anderson impurity model, see
  Ref.~\onlinecite{Karlis08}).  The above choice of $k(t)$ requires
  the implementation of Monte Carlo updates which change the
  perturbation order from $n$ to $n\pm 2$. We found, however, that the
  odd perturbation orders are continuously suppressed as $k(t)$
  approaches $- U/4$, so one may as well choose $k(t)=-U/4+\delta$
  (with small $\delta$) in combination with rank one updates.

  The efficiency of the Green function measurement can be improved
  dramatically by the following simple tricks. First, we rewrite
  Eq.~(\ref{tildeG}) as
  \begin{widetext}
    \begin{equation}
      G^c_\sigma(t,t')=G_{0,\sigma}(t,t')+\int_\mathcal{C} \!\! ds_1 \! \int_\mathcal{C} \!\! ds_2 G_{0,\sigma}
      (t,s_1) \Big \langle i \sum_{i,j=1}^n \delta_\mathcal{C}(s_1,t_i)[(e^{\Gamma_
        \sigma}-1)N_\sigma]_{i,j}\delta_\mathcal{C}(s_2,t_j)\Big\rangle_{mc} G_{0,\sigma}(s_2,
      t'),
      \label{X}
    \end{equation}
  \end{widetext}
  where the variables $s_1$ and $s_2$ run over the contour
  $\mathcal{C}$, 
  and $\expval{ \cdot }_{mc}$ denotes the Monte Carlo averaging. 
  It is therefore sufficient to accumulate the 
  impurity system $T$-matrix
  \begin{equation}
    X_\sigma(s_1, s_2)=\Big\langle i \sum_{i,j=1}^n \delta_\mathcal{C}(s_1,t_i)
    [(e^{\Gamma_\sigma}-1)N_\sigma]_{i,j}\delta_\mathcal{C}(s_2,t_j)\Big\rangle_{mc},
    \label{Xonly}
  \end{equation}
  as mentioned in Ref.~\onlinecite{Gull08}. While the measurement of
  $X$ on some fine grid introduces discretization errors, these can be
  made negligibly small at essentially no computational cost.
  Furthermore, comparison of Eq.~(\ref{X}) to the Dyson  equation 
  (\ref{dmft-dysonloc}) shows
  that $X$ is related to the self-energy by
  \begin{equation}
    \label{qmc-x}
    X \convz G_0=\Sigma \convz G,
  \end{equation}
  so the measurement of $X$ allows to extract $\Sigma$ as explained in
  Section~\ref{sec-qmcdmft} C.

  Further improvements are possible. Assuming that the perturbation
  order on the real-time branch is non-zero, it follows from
  Eq.~(\ref{weight}) that the weight of the Monte Carlo configuration
  changes sign if the last spin (corresponding to the largest time
  argument) is shifted from the forward contour to the backward
  contour or vice versa. Since the absolute value of the weight does
  not change, these two configurations will be generated with equal
  probability. As a result, all terms in Eq.~(\ref{Xonly}) which
  do not involve the last operator on the contour will cancel on average. 
  It is therefore more efficient to accumulate only the
  contributions to Eq.~(\ref{Xonly}) from those pairs $(i,j)$ in which
  either $i$ or $j$ corresponds to the last operator on the real-time
  branch.  (If all spins sit on the imaginary-time branch, no such
  simplification is possible.)  We also note that the error bars on
  measurements can be substantially reduced by appropriate
  symmetrizations of the real and imaginary parts of $X$ (symmetry
  lines $s_1=t_\text{max}$, $s_2=t_\text{max}$, $s_1=s_2$).

  \section{Weak-coupling CTQMC + DMFT}
  \label{sec-qmcdmft}

  To use the weak-coupling CTQMC as an impurity solver within DMFT, we
  iterate the following two steps until convergence: (i) The local
  Green function $G_\sigma(t,t')$ is determined in CTQMC
  [Eq.~(\ref{X})], using the noninteracting bath Green function
  $G_{0,\sigma}(t,t')$ as input, and (ii), $G_{0,\sigma}(t,t')$ is
  determined from its inverse (\ref{dmft-g0loc-inv}), using the QMC
  output $G_\sigma(t,t')$ and the self-consistency
  Eq.~(\ref{selfconsistency-bethe}). We start the iteration from an
  initial guess for $G_{0,\sigma}(t,t')$, for which we usually take
  the noninteracting equilibrium Green function,
  \begin{equation}
    \label{dmft-guess}
    G_{\sigma}^\text{eq}(t,t')
    =
    i
    \int \!d\epsilon \,\rho(\epsilon)
    e^{i\epsilon(t'-t)} [f(\epsilon)-\Theta_\mathcal{C}(t,t')],
  \end{equation}
  where $\Theta_\mathcal{C}(t,t')=1$ if $t$ is later on the contour
  than $t'$ and otherwise zero.  
  
  In this section we describe in detail how $G_{0,\sigma}$ is
  determined from $\Lambda_\sigma$ (Sec.~\ref{sec-cinv}), how the
  self-energy is calculated from the impurity correlation function
  $X_\sigma$ 
  after convergence of the DMFT iteration
  (Sec.~\ref{sec-self-energy}), and how one finally obtains
  expectation values of various observables of the lattice system
  (Sec.~\ref{sec-observables}).  Furthermore, we introduce a real
  frequency representation which is needed to efficiently treat the
  case of zero temperature on the L-shaped contour
  (Sec.~\ref{sec-real-frequency}), and combine this with the
  weak-coupling impurity solver for the case of a noninteracting
  initial state.

  \subsection{Integral equations on the contour $\CC$}
  \label{sec-cinv}

  \renewcommand{\tmin}{0}

  Within nonequilibrium DMFT  one must frequently solve 
  equations on $\CC$ of the type
  \begin{align}
    \label{ky-cinv}
    \big[ i\partial_t  -h(t) \big] \YY(t,t')
    -[\KK \convz \YY ] (t,t')
    &=
    \delC(t,t')
  \end{align}
  with a known integral kernel $\KK(t,t')$.  The solution $\YY(t,t')$
  is unique when the antiperiodic boundary condition (\ref{ky-bc}) is
  imposed on $\YY(t,t')$. For example, both Eq.~(\ref{pseudodyson})
  for the noninteracting bath Green function $G_{0,\sigma}(t,t')$ and
  Eq.~(\ref{dmft-lattice-dyson-kspace}) for the momentum-resolved
  Green function have this form.

  By choosing a suitable discretization of the contour $\CC$,
  Eq.~(\ref{ky-cinv}) can in principle be reduced to the inversion of
  a matrix whose dimension is given by the number of mesh points along
  $\CC$.\cite{Freericks2008b} In the following we pursue a different
  approach, where both $\YY$ and $\KK$ in Eq.~(\ref{ky-cinv}) are
  first represented in terms of their respective real and imaginary
  time components (\ref{ky-cntr-mat})-(\ref{ky-cntr-mix}), and
  separate integral equations (which are similar to the Kadanoff-Baym equations\cite{keldyshintro}) are solved for each component.
  Although this procedure may seem rather cumbersome compared to direct contour
  discretization, it has several advantages: (i) It is straightforward
  to incorporate the hermitian symmetry (\ref{ky-cntr-sym}) which is
  satisfied by both the local self-energy and the hybridization function
  $\Lambda_\sigma$. (ii) The resulting equations are
  Volterra type integro-differential equations, for which highly
  stable and accurate algorithms can be found in the
  literature,\cite{Brunner1986a} and which remain causal even when
  they are approximated numerically. Finally, (iii), the
  real-frequency representation which we introduce in
  Sec.~\ref{sec-real-frequency} to handle initial states at zero
  temperature is based on this approach.

  In the following we assume that $\YY$ and $\KK$ satisfy the
  hermitian symmetry (\ref{ky-cntr-sym}), such that it is sufficient
  to determine the Matsubara, retarded, mixed ``$\tv$'', and lesser
  components of $\YY$. Corresponding components of the convolution
  $\KK \convz \YY$ in Eq.~(\ref{ky-cinv}) are obtained from the
  Langreth rules,\cite{keldyshintro} which follow directly from the
  definitions (\ref{ky-cntr-mat})-(\ref{ky-cntr-mix}) and the
  definition of the contour integral.  By taking the Matsubara
  component (\ref{ky-cntr-mat}) of Eq.~(\ref{ky-cinv}) we obtain
  \begin{multline}
    \label{ky-cinv-mat-eom}
    (-\partial_\tau -h)
    \YY^\mat(\tau,\tau')
    +i\int\limits_0^\beta d\bar \tau \KK^\mat(\tau,\bar\tau)\YY^\mat(\bar \tau,\tau')
    =
    \\
    i\delta(\tau-\tau'),
  \end{multline}
  where $h=h(0)$ is constant on the imaginary branch. This equation
  must be augmented with an antiperiodic boundary condition
  $\YY^\mat(0,\tau')$ $=$ $-\YY^\mat(\beta,\tau')$ which follows from
  Eq.~(\ref{ky-bc}).  When we assume that the kernel $\KK^\mat$ has the
  Matsubara frequency representation (\ref{ky-mat}), it follows
  that the solution $\YY^\mat(\tau,\tau')$ is of the same form, with
  \begin{align}
    \label{ky-cinv-mat-w}
    y^\mat(i\omega_n) = [i\omega_n - h -k^\mat(i\omega)]^{-1}.
  \end{align}
  As required by causality, $\YY^\mat$ thus  turns out to depend only on the initial 
  equilibrium state, independent of the subsequent perturbation of  the system.

  In a similar fashion, the retarded component (\ref{ky-cntr-ret})  of 
  Eq.~(\ref{ky-cinv}) is given by 
  \begin{align}
    \label{ky-cinv-ret-eom}
    \big[i\partial_t -h(t)\big]
    \YY^\ret(t,t')
    -\int\limits_{t'}^t \!d\tb \,\KK^\ret(t,\tb)\YY^\ret(\tb,t')
    =\delta(t-t').
  \end{align}
  Because $\YY^\ret(t,t')$ vanishes for $t<t'$ by definition [cf.~Eq.~(\ref{ky-cntr-ret1})],
  integration over the $\delta$-function yields 
  \begin{equation}
    \label{grettt}
    \YY^\ret(t,t) =-i.
  \end{equation}
  One can thus restrict the solution of Eq.~(\ref{ky-cinv-ret-eom}) to
  $t>t'$, drop the $\delta$-function on the right-hand side and
  instead impose (\ref{grettt}) as an initial condition.

  The limits of the integral in (\ref{ky-cinv-ret-eom}) take into
  account that retarded functions vanish for $t>t'$. This fact
  turns Eq.~(\ref{ky-cinv-ret-eom}) into a Volterra equation
  of second kind,\cite{Brunner1986a} i.e., the derivative at time $t$
  is determined by the kernel and the function at earlier times only.
  The numerical solution of this type of equations is analogous to the
  solution of ordinary differential equations.\cite{Brunner1986a}

  For the Green functions (\ref{ky-cntr-mix}) with mixed time 
  arguments, Eq.~(\ref{ky-cinv}) reads
  \begin{multline}
    \label{ky-cinv-tv-eom}
    [i\partial_t -h(t)]
    \YY^\tv(t,\tau)
    -
    \int\limits_{\tmin}^t\! d\tb\, \KK^\ret(t,\tb)\YY^\tv(\tb,\tau)
    \\
    =
    -i\int\limits_0^\beta \!d\bar\tau\, \KK^\tv(t,\bar\tau)\YY^\mat(\bar\tau,\tau)\,.
  \end{multline} 
We assume that $\YY$ is continuous on $\CC$
(which is true if neither $\KK(t,t')$ nor $h(t)$ are singular at $t=0$), 
such that Eq.~(\ref{ky-cinv-tv-eom}) 
must be solved with the initial condition
  $\YY^\tv(0,\tau)=\YY^\mat(0,\tau)$. For given $\tau$, Eq.~(\ref{ky-cinv-tv-eom})
  is an inhomogeneous Volterra integro-differential equation, for
  which only known  functions [cf.~Eq.~(\ref{ky-cinv-mat-w})] enter 
  the source term on the right-hand side.

  A third and last Volterra integral equation can be derived for the
  lesser component (\ref{ky-cntr-les}),
  \begin{multline}
    \label{ky-cinv-les-eom}
    \big[i\partial_t -h(t)\big]
    \YY^\les(t,t')
    -\int\limits_{\tmin}^t\! d\tb \,\KK^\ret(t,\tb)\YY^\les(\tb,t')
    =
    \\
    -i\!\int\limits_0^\beta \!\!d\bar\tau \KK^\tv(t,\bar\tau)\YY^\vt(\bar\tau,t')
    +\! \int\limits_0^{t'}\!\! d\tb \KK^\les(t,\tb)\YY^\adv(\tb,t').\!\!\!
  \end{multline}
  Due to the symmetry (\ref{ky-cntr-sym}) it is sufficient to solve
  this equation for $t<t'$, with the initial condition
  $\YY^\les(\tmin,t')$ $=$ $-\YY^\vt(\beta,t')$. The latter follows
  from Eq.~(\ref{ky-bc}) and the continuity of $\YY$ along $\CC$.  The
  functions $\YY^\adv$ and $\YY^\vt$ which enter the source term of
  Eq.~(\ref{ky-cinv-les-eom}) on the right-hand side can be obtained from
  the previous solution of Eqs.~(\ref{ky-cinv-ret-eom}) and
  (\ref{ky-cinv-tv-eom}), and the symmetry (\ref{ky-cntr-sym}).
  The successive solution of Eqs.~(\ref{ky-cinv-mat-eom}),
  (\ref{ky-cinv-ret-eom}), (\ref{ky-cinv-tv-eom}), and
  (\ref{ky-cinv-les-eom}) completes the determination of the contour
  function $\YY$.

  \subsection{Determination of the self-energy}
  \label{sec-self-energy}

  The impurity self-energy can be obtained from the correlation
  function $X_\sigma$ via Eq.~(\ref{qmc-x}). By comparison of the
  Dyson equation (\ref{dmft-dysonloc}) in integral form, $G_{\sigma}$
  $=$ $G_{0,\sigma}$ $+$ $G_{\sigma} \convz \Sigma_{\sigma} \convz
  G_{0,\sigma}$, with Eq.~(\ref{X}), i.e.,
 $G_{\sigma}$
  $=$ $G_{0,\sigma}$ $+$ $G_{0,\sigma} \convz X_{\sigma} \convz
  G_{0,\sigma}$, we find the relation
  \begin{equation}
    \label{dmft-getsigma}
    (1 + X_{\sigma} \convz G_{0,\sigma}) \convz \Sigma_{\sigma}
    =
    X_{\sigma}.
  \end{equation}
  This equation is very similar to Eq.~(\ref{ky-cinv}), with unknown 
  $\YY=\Sigma$, kernel $\KK=X_{\sigma} \convz G_{0,\sigma}$, $h=1$,
  and without the differential term.  The solution of (\ref{dmft-getsigma}) is thus analogous to 
  Eq.~(\ref{ky-cinv}), using a decomposition in terms of the components 
  (\ref{ky-cntr-mat})-(\ref{ky-cntr-les}). The final equations read
  \begin{subequations}
    \label{ky-sigma-eom}
    \begin{align}
      \label{ky-sigma-mat-eom}
      \Sigma^\mat&(t\omega_n)
      =
      \frac{x^\mat(i\omega_n)}{1+k^\mat(i\omega_n)},
      \\
      \label{ky-sigma-ret-eom}
      \Sigma^\ret&(t,t')
      +\int\limits_{t'}^t \!d\tb \,\KK^\ret(t,\tb)\Sigma^\ret(\tb,t')
      =
      X^\ret(t,t'),
      \\
      \label{ky-sigma-tv-eom}
      \Sigma^\tv&(t,\tau)
      +\int\limits_{0}^t \!d\tb \,\KK^\ret(t,\tb)\Sigma^\tv(\tb,\tau)
      =X^\tv(t,\tau) +
      \nonumber\\
      &i\int\limits_{0}^\beta \!d\bar\tau \,\KK^\tv(t,\bar\tau)\Sigma^\mat(\bar\tau,\tau)
      \\
      \label{ky-sigma-les-eom}
      \Sigma^\les&(t,t')
      + \int\limits_{\tmin}^t\! d\tb \,\KK^\ret(t,\tb)\Sigma^\les(\tb,t')
      =X_\sigma^\les(t,t')+
      \nonumber\\
      &i\!\int\limits_0^\beta \!\!d\bar\tau \KK^\tv(t,\bar\tau)\Sigma^\vt\!(\bar\tau,t')
      \!-\!\!\!
      \int\limits_0^{t'}\!\!d\tb\, \KK^\les(t,\tb)\Sigma^\adv\!(\tb,t').
    \end{align}
  \end{subequations}
  Note that the kernel $\KK=X_{\sigma}\convz G_{0,\sigma}$ does not satisfy 
  the hermitian symmetry (\ref{ky-cntr-sym}), i.e., $X_{\sigma}\convz G_{0,\sigma}$
  $\neq$ $G_{0,\sigma}\convz X_{\sigma}$.

  We would like to remark that the self-energy can equally well be 
  determined from the linear equation
  \begin{equation}
    \label{sigmag}
    X_{\sigma} \convz G_{\sigma0} = \Sigma_{\sigma} \convz G_{\sigma}.
  \end{equation}
  However, Eqs.~(\ref{ky-sigma-eom}) are essentially Volterra integral
  equations of the second kind, while Eq.~(\ref{sigmag}) leads to
  Volterra equations of the first kind, i.e., only the integral-term
  is present on the left-hand side.  Because the numerical solution of
  Volterra equations of the first kind tends to be
  unstable\cite{Brunner1986a} we prefer the solution of Eq.~(\ref{dmft-getsigma}) 
  over Eq.~(\ref{sigmag}).

  \subsection{Expectation values of observables}
  \label{sec-observables}

  From the self-energy $\Sigma$ one can directly compute the
  expectation values of observables of the lattice Hamiltonian. In
  this section we let $\expval{\cdots}$ denote the initial state
  expectation value at temperature $T=1/\beta$, and operators with hat
  are in Heisenberg representation with respect to the Hubbard
  Hamiltonian (\ref{hubbard}) with time-dependent interaction. The
  number of lattice sites will be denoted by $L$.

  The particle number per site for spin $\sigma$ is given by the
  local Green function $G_\sigma(t,t')$
  \begin{equation}
    \label{density}
    n_\sigma(t) 
    =
    \frac{1}{L}\sum_j
    \expval{\hat c_{j\sigma}^\dagger(t) \hat c_{j\sigma}(t)} 
    =
    -i G_{\sigma}^\les(t,t),
  \end{equation}
  provided that the state is homogeneous.
  Because $n_\sigma(t)$ is conserved, the condition $G_{\sigma}^\les(t,t)= \text{const}$ 
  provides a first test of the numerical accuracy.

  The occupation of the  momentum states
  \begin{equation}
    \label{neps}
    n(\epsilon_\Vk,t) \equiv
    \expval{\hat c_\Ks^\dagger(t) \hat c_\Ks(t)} 
    =
    -i G_\Ks^\les(t,t),
  \end{equation}
  is obtained from the momentum-resolved Green function $G_\Ks(t,t')$
  $=$ $-i\expval{\TC \hat c_\Ks(t) c_\Ks^\dagger(t')}$. For a
  momentum independent $\Sigma_\sigma$, $n(\epsilon_\Vk,t)$ depends on
  momentum $\Vk$ only via the band-energy $\epsilon_\Vk$. The Green
  function $G_\Ks(t,t')$ is determined from the lattice Dyson equation
  (\ref{dmft-lattice-dyson-kspace}), whose solution is analogous to that 
  of  Eq.~(\ref{ky-cinv}).  The kinetic energy per lattice site
  \begin{subequations}
    \label{ekin}
    \begin{equation}
      E_\text{kin}(t)
      =
      \frac{1}{L} \sum_{\Ks} \epsilon_\Vk \expval{\hat c_\Ks^\dagger(t) \hat c_\Vk(t)},
    \end{equation}
    is obtained from $n(\epsilon,t)$ by replacing the $\Vk$-sum 
    with an integral over the local density of states [Eq.~(\ref{rho-semi})], 
    \begin{equation}
      E_\text{kin}(t) = 
      \int\!d\epsilon\,\rho(\epsilon)\,\epsilon \,n(\epsilon,t).
    \end{equation}
  \end{subequations}

  Furthermore we are interested in the double occupation per lattice site
  \begin{equation}
    \label{docc}
    d(t)  = \frac{1}{L}
    \sum_i \expval{\hat n_{i\uparrow}(t) \hat n_{i\downarrow}(t) },
  \end{equation}
  and the interaction energy
  \begin{align}
    \label{epot}
    E_\text{pot} 
    &\equiv 
    U(t) \sum_i \big\langle
    \big(\hat n_{i\uparrow}(t)-\tfrac12\big)
    \big(\hat n_{i\downarrow}(t)-\tfrac12\big)
    \big\rangle
    \\
    &=
    U(t)\big[d(t)-\tfrac12(n_{\uparrow}(t)+n_{\downarrow}(t))+\tfrac14\big].
  \end{align}
  To calculate this quantity we consider the equation of motion for 
  the local lattice Green function $G_{jj\sigma}$, which reads
  \begin{align}
    &[(G_{\sigma}^{-1})_{jl} 
    \convz G_{lj\sigma}]
    (t,t')
    =
    \delC(t,t') + U(t)\Gamma_{j\sigma}(t,t'),
    \\
    &(G_{\sigma}^{-1})_{jl}(t,t')
    =
    \delC(t,t')[\delta_{jl}(i\partial_t + \mu) - t_{jl}],
    \\
    \label{gamma-lattice}
    &\Gamma_{j\sigma}(t,t')
    =
    -i\expval{\TC \hat c_{i\sigma} (t) (\hat n_{i\bar\sigma}(t)-\tfrac12) \hat c_{i\sigma}^\dagger(t')}.
  \end{align}
  Comparison with the lattice Dyson equation in real space yields
  \begin{equation}
    U(t)\Gamma_{j\sigma}(t,t')
    =
    [\Sigma_{\sigma} \convz G_{jj\sigma}](t,t'),
  \end{equation}
  because the self-energy is local and site-independent.  Hence 
  $\Gamma_{\sigma}\equiv\Gamma_{j\sigma}$ can be determined
  from quantities measured in the single-site problem 
  [cf.~Eq.~(\ref{sigmag})], and Eq.~(\ref{gamma-lattice}) 
  implies 
  \begin{equation}
    d(t) = -i \Gamma_{i\sigma}^\les(t,t) +\tfrac12 n_{\sigma}(t)
  \end{equation}
 for a homogeneous state.
  
  Finally we can compute the total energy
  from Eqs.~(\ref{ekin}) and (\ref{epot}),
  \begin{equation}
    \label{etot}
    E_\text{tot}(t) = E_\text{kin}(t) + E_\text{pot}(t). 
  \end{equation}
  This quantity must be constant when the Hamiltonian is time-independent,
  which provides a second test for the accuracy of the numerical
  solution.

  \subsection{Real-frequency representation}
  \label{sec-real-frequency}

  In this subsection we introduce a partial Fourier transform of the
  mixed components ``$\tv$'' and ``$\vt$'', which will allow us to
  handle contour equations such as Eqs.~(\ref{ky-cinv}) and
  (\ref{dmft-getsigma}) in the limit of zero temperature without
  dealing explicitly with a contour of infinite length.

  We start from the Fourier series in the interval 
  $0\le\tau\le \beta$,
  \begin{subequations}
    \label{ky-mixed-ft}
    \begin{align}
      \label{ky-vt-ft1}
      \YY^\vt(i\omega_n,t) &= \int\limits_0^\beta \!d\tau \,\YY^\vt(\tau,t) e^{i\omega_n\tau},
      \\
      \label{ky-vt-ft2}
      \YY^\vt(\tau,t) &= \frac{1}{\beta} \sum_n \YY^\vt(i\omega_n,t) e^{-i\omega_n\tau},
    \end{align}
  \end{subequations}
  in terms of fermionic Matsubara frequencies $i\omega_n$. 
  This representation is now used within the solution
  of Eq.~(\ref{ky-cinv}). In contrast to Eq.~(\ref{ky-cinv-tv-eom}), 
  the corresponding equation for the mixed ``$\vt$'' component
  \begin{multline}
    \label{ky-cinv-vt-eom}
    (-\partial_\tau -h)
    \YY^\vt(\tau,t)
    +i\int\limits_0^\beta \!d\bar\tau\, \KK^\mat(\tau,\bar\tau)\YY^\vt(\bar\tau,t)
    \\
    =
    \int\limits_{\tmin}^t\! d\tb\, \KK^\vt(\tau,\tb)\YY^\adv(\tb,t)
  \end{multline}
  is not an initial value problem,
  but rather a boundary value problem on $\CC$: 
  The boundary condition $\YY^\vt(\beta,t)$ $=$ $-\YY^\vt(0,t) - \YY^\adv(0,t)$
  follows from Eqs.~(\ref{ky-bc}) and (\ref{ky-cntr-adv}) and the continuity 
 of the  contour functions 
  along $\CC$.  Using the transformation (\ref{ky-mixed-ft}) and Eq.~(\ref{ky-cinv-mat-w})
  for the Matsubara component,  Eq.~(\ref{ky-cinv-vt-eom}) becomes an 
 explicit integral expression  for $\YY^\vt(i\omega_n,t)$,
  \begin{multline}
    \label{ky-cinvc-tv-w}
    \YY^\vt(i\omega_n,t)
    =
    \\
    y^\mat(i\omega_n)
    \Big[
    \YY^\adv(0,t)
    +
    \int\limits_{\tmin}^t\!\! d\tb\, \KK^\vt(i\omega_n,\tb) \YY^\adv(\tb,t) \Big].
  \end{multline}

  In the following we assume that $\KK^\vt(i\omega_n,t)$ can be
  continued to complex frequencies $z$, such that $\KK^\vt(z,t)$ is
  analytic in the upper and lower complex half plane, respectively,
  and has a branch cut along the real frequency axis. For the Green
  function (\ref{ky-cntr}) this property follows from a Lehmann
  representation in terms of an eigenbasis $\{\ket{n}\}$ of $H(0)$,
  \begin{equation}
    G^\vt\sss(z,t)
    \!=
    i
    \sum_{nm} 
    \frac{(w_n\!+\!w_m)\expval{n| c_\alpha|m}  \expval{m|c_{\alpha'}^\dagger(t)|n}}
    {z+E_n-E_m},
  \end{equation}
  which has poles on the real axis only ($w_n=e^{-\beta E_n}$).

  When Eq.~(\ref{ky-cinvc-tv-w}) is continued to the real axis, we
  obtain two functions $\YY^\vt(\omega^\pm,t)$, where
  $\omega^\pm=\omega\pm i\eta$ for $\eta\to 0^+$. In contrast to the
  equilibrium functions $y^\mat(\omega^\pm)$, the two functions
  $\YY^\vt(\omega^\pm,t)$ are not simply related by complex
  conjugation. Matsubara summations are then transformed into
  integrals along the branch cut of $\YY^\vt(z,t)$ in the usual way.
  For example, the backtransformation (\ref{ky-vt-ft2}) is given by
  \begin{align}
    \YY^\vt(\tau,t)
    =\!\!
    \int \frac{d\omega}{2\pi i} f(\omega) e^{\omega\tau}
    [\YY^\vt(\omega^-\!,t)- \YY^\vt(\omega^+\!,t)],
  \end{align}
  where $f(\omega)=1/(e^{-\beta\omega}+1)$ is the Fermi function.

  Using the real frequency representation for the mixed components, the
  first source term on the right hand side of
  Eq.~(\ref{ky-cinv-les-eom}) can be rewritten in terms of the
  previously determined functions $\YY^\vt(\omega^\pm,t)$
  \begin{align}
    \label{ky-cinv-source-w}
    &\int\limits_0^\beta \!d\bar\tau\, \KK^\tv(t,\bar\tau)\YY^\vt(\bar\tau,t)
    =
    \int \!\frac{d\omega}{2\pi i}\, f(\omega) 
    \,\,\times
    \nonumber\\
    &~~\big[
    \KK^\tv(t,\omega^+)\YY^\vt(\omega^+,t')
    -\KK^\tv(t,\omega^-)\YY^\vt(\omega^-,t')
    \big].
  \end{align}
  Here the Fourier transformation (\ref{ky-vt-ft1}) with opposite sign 
  for $i\omega_n$ is used in the second time argument for the 
  $\tv$ Green functions  (\ref{ky-cntr-tv}), such that
  \begin{equation}
    \label{ky-cntr-vtzsym}
    \YY^\vt(z,t) = -\YY^\tv(t,z^*)^*
  \end{equation}
  follows by symmetry (\ref{ky-cntr-sym-mix}). 

  The above real-frequency representation can be used within DMFT
  whenever the impurity problem is solvable at zero temperature. This
  is the case for approximate analytical methods (Sec.
  \ref{dmft-sec-ipt}). It might also be of advantage for arbitrary
  initial states in the Falicov-Kimball
  model,\cite{Freericks2008b,Eckstein2008a} where the solution of the
  impurity is based on the solution of equations of motion which have
  exactly the structure of Eq.~(\ref{ky-cinv}). In the following
  section we present another application, namely nonequilibrium DMFT
  for the Hubbard model with a noninteracting initial state.

  \subsection{CTQMC and DMFT for noninteracting initial states}
  \label{sec-flowdiagram}

  \begin{figure}
    \centerline{\includegraphics[width=1.0\columnwidth,bb = 71 642 321 738,clip=true]{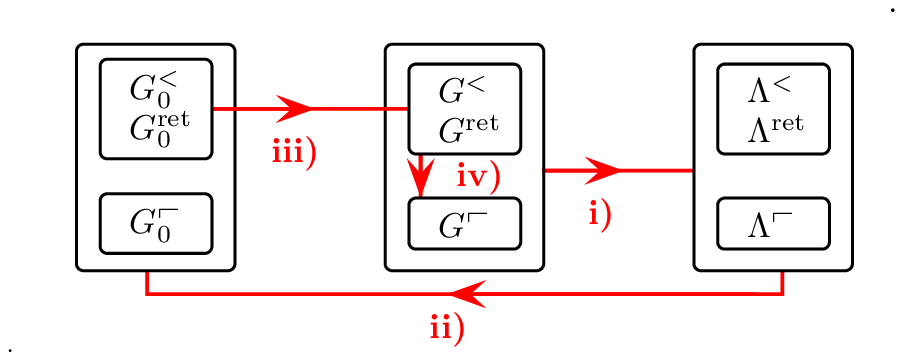}}
    \caption{
      Computational scheme for the nonequilibrium DMFT 
      using the self-consistency (\ref{selfconsistency-bethe}) 
      and noninteracting initial states. Steps (i)-(iv) are explained
      in the text.}
    \label{dmft-fig-calc}
  \end{figure}

  The computational scheme for the solution of the nonequilibrium DMFT
  equations is represented in Fig.~\ref{dmft-fig-calc} for a
  semielliptic density of states (\ref{rho-semi}) and a noninteracting
  initial state.  Green functions $G_\sigma$, $G_{\sigma0}$, and
  $\Lambda_\sigma$ satisfy the symmetry (\ref{ky-cntr-sym}), such that
  they are represented by their Matsubara, retarded, ``$\vt$'' and
  lesser component. The Matsubara Green functions are given by the
  equilibrium (noninteracting) Green function
  \begin{align}
    \label{dmft-gmat}
    g_{\sigma}^\mat(i\omega_n) 
  &= g_{0,\sigma}^\mat(i\omega_n) =
    \int \!d\epsilon\, 
    \frac{\rho(\epsilon)}{i\omega_n + \mu -\epsilon} 
    \\
  &=
    V^{-2} \lambda_\sigma^\mat(i\omega_n),
  \end{align}
  where $\rho(\epsilon)$ is given by Eq.~(\ref{rho-semi}), and the
  last equality holds due to the self-consistency
  (\ref{selfconsistency-bethe}).  The mixed components $G_\sigma^\vt$,
  $G_{\sigma0}^\vt$, and $\Lambda_\sigma^\vt$ are represented after
  the partial Fourier transform (\ref{ky-mixed-ft}) and analytical
  continuation by their value along the branch cut, i.e., for each
  Green function $\YY$ we keep two functions $\YY^\vt(\omega^\pm,t)$
  on a fixed frequency mesh.

  The DMFT iteration is started from an initial guess
  (\ref{dmft-guess}). In step (i) [cf. Fig. \ref{dmft-fig-calc}], the
  Weiss field $\Lambda(t,t')$ is computed from the closed
  self-consistency equation (\ref{selfconsistency-bethe}). This is
  then used to determine the noninteracting bath Green function
  $G_{\sigma0}$ from its inverse (\ref{dmft-g0loc-inv}), as explained
  in the previous subsection [step (ii) in Fig. \ref{dmft-fig-calc}].
  The function $G_{\sigma0}$ is the input for the calculation of the
  interacting bath Green function (\ref{dmft-gloc}) using CTQMC [step
  (iii) in Fig. \ref{dmft-fig-calc}]. Because the initial state is
  noninteracting, the Monte Carlo simulation is restricted to the
  real-time branch of the contour, and  only the real-time
  components $G_\sigma^\ret$ and $G_\sigma^\les$ are obtained.
  However, the mixed component $G_\sigma^\vt(\omega^\pm,t)$ can be
  reconstructed from these functions and the previous Weiss field
  $\Lambda$ [step (iv)]: For this purpose consider the Dyson equation
  (\ref{dmft-dysonloc}), which has the form of Eq.~(\ref{ky-cinv})
  after the replacement $\KK=\Lambda_\sigma+\Sigma_\sigma$,
  $\YY=G_\sigma$, and $h(t)=\mu$. Hence $G^\vt(z,t)$ can be obtained
  from the integral (\ref{ky-cinvc-tv-w}), making the same
  replacements. Because $\Sigma_\sigma(t,t')$ is
  proportional to the interaction strengths $U(t)$ and $U(t')$, we
  have $\Sigma_\sigma^\vt(\tau,t)=0$ for a noninteracting initial
  state, with $U(-i\tau)=U(0)=0$. Hence $G^\vt(\omega^\pm,t)$ is given
  by
  \begin{multline}
    G_\sigma^\vt(\omega^\pm,t)
    \\
    =
    g_\sigma^\mat(\omega^\pm)
    \Big[
    G_\sigma^\adv(0,t)
    +
    \int\limits_{\tmin}^t\!\! d\tb\, \Lambda_\sigma^\vt(\omega^\pm,\tb) G^\adv_\sigma(\tb,t) \Big],
  \end{multline}
  where $g_\sigma^\mat(\omega^\pm)$ $=$ $\mp \pi i \rho(\omega)$
  [Eq.~(\ref{dmft-gmat})]. Steps (i) through (iv) are repeated until
  convergence, which is usually reached after not more that $15$
  iterations.

  \section{Comparison to iterated perturbation theory}
  \label{dmft-sec-ipt}

  \begin{figure}
    \begin{center}
      \includegraphics[width=\columnwidth]{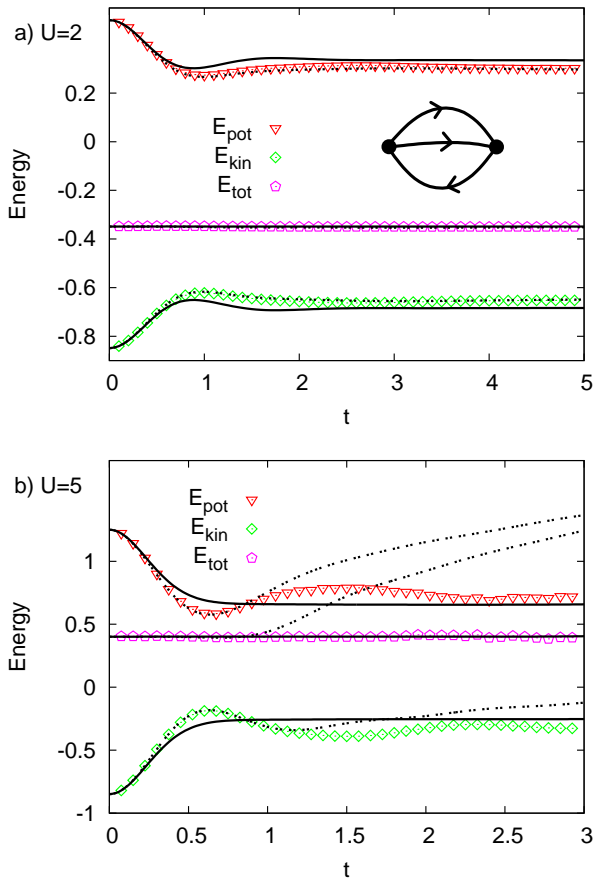}\\
    \end{center}\caption{
      Potential energy $E_\text{pot}(t)$
      [Eq.~(\ref{epot})], kinetic energy $E_\text{kin}(t)$
      [Eq.~(\ref{ekin})], and total energy $E_\text{tot} =
      E_\text{pot}+E_\text{kin}$ in the half-filled Hubbard model
      after an interaction quench from $U=0$ to $U=2$ (a)
      and $U=5$ (b). The initial state temperature is given
      by $T$ $=$ $0$. The results were obtained with nonequilibrium
      DMFT for a semielliptic density of states (\ref{rho-semi}),
      using either CTQMC (symbols), IPT (dashed lines) or SPT (solid
      lines) to solve the single-site problem. The inset in the upper
      panel shows the second-oder diagram for $\Sigma_\sigma$. Lines
      represent $G_{0,\sigma}$ for IPT [Eq.~(\ref{dmft-ipt})], and
      $G_\sigma$ for SPT [Eq.~(\ref{dmft-spt})].
    }
    \label{dmft-fig-energy}
  \end{figure}

  In equilibrium DMFT, the so-called iterated perturbation theory
  (IPT),\cite{Zhang1993a,Georges1996} is frequently used as an
  approximate but efficient method to solve the single-site
  problem. Within IPT, the self-energy $\Sigma_\sigma$ is expanded up
  to second order in the interaction $U$.
  Although this is a weak-coupling expansion
  by construction, it is accidentally correct for the atomic limit of
  the half-filled Hubbard model in equilibrium. In
  many aspects, IPT thus provides a reasonable interpolation between
  the two exact limits $U=0$ and $V=0$. In particular, it
  qualitatively reproduces the DMFT phase diagram and the Mott transition 
  in the paramagnetic phase, although there are quantitative differences
  to numerically exact QMC results. It is therefore interesting to see
  whether this approximation performs similarly well when used to
  solve the single-site problem in nonequilibrium DMFT.

  In the following we restrict ourselves to the half-filled Hubbard
  model with time-dependent interaction $U(t)$. The Hartree
  contribution to the self-energy (first-order diagram), which gives
  a shift of the chemical potential with respect to $\mu$ $=$ $0$,
  then vanishes, and the second-order contribution to the self-energy 
  is given by a single diagram (inset in Fig.~\ref{dmft-fig-energy}a),
  \begin{multline}
    \label{dmft-ipt}
    \Sigma^{\text{ipt}}_\sigma(t,t') = 
    - U(t) U(t') G_{0,\sigma}(t,t') G_{0,\bar\sigma}(t',t) G_{0,\bar\sigma}(t,t').
  \end{multline}
  This equation is easily incorporated into the DMFT self-consistency
  iteration by replacing step (iii) and (iv) in
  Fig.~\ref{dmft-fig-calc} with a solution of the Dyson equation
  (\ref{dmft-dysonloc}) for $G_\sigma$, where $\Sigma_\sigma$ is given
  by Eq.~(\ref{dmft-ipt}).  Equation (\ref{dmft-dysonloc}) is solved
  numerically, as described in Sec.~\ref{sec-cinv}.

  In Fig.~\ref{dmft-fig-energy} we plot the potential energy
  $E_\text{pot}(t)$ [Eq.~(\ref{epot})], the kinetic energy
  $E_\text{kin}(t)$ [Eq.~(\ref{ekin})], and total energy $E_\text{tot}
  = E_\text{pot}+E_\text{kin}$ of the half-filled Hubbard model after
  an interaction quench from the noninteracting initial state at
  temperature $T=0$. The hopping matrix elements correspond to
  a semielliptic density of states Eq.~(\ref{rho-semi}) with quarter 
  bandwidth $V=1$, and time is measured in
  units of $\hbar/V=1$. The numerically exact CTQMC results show a
  rapid relaxation of these quantities, which is discussed in detail
  below.  As required by energy conservation, $E_\text{tot}$ is
  constant within the numerical accuracy. IPT can reproduce these
  results rather accurately for small values of $U$
  (Fig.~\ref{dmft-fig-energy}a). Already at intermediate coupling,
  however, the results of CTQMC and IPT strongly deviate from each
  other (Fig.~\ref{dmft-fig-energy}b).  In particular, the total
  energy $E_\text{tot}$ is generally not conserved within IPT, such
  that the use of IPT as an approximation for the intermediate- and
  strong-coupling regime becomes highly questionable. In contrast to
  equilibrium DMFT, IPT does not provide a reasonable interpolation
  between weak- and strong-coupling regimes.

  This violation of energy conservation is cured by a simple procedure.
  An expansion of $\Sigma_\sigma$ up to finite order in terms of the
  noninteracting Green function is not a conserving approximation in
  the sense of Kadanoff and Baym.\cite{Baym1961a,Baym1962a} However,
  the approximation becomes conserving when $G_{0,\sigma}$ in
  Eq.~(\ref{dmft-ipt}) is replaced by the full interacting Green
  function,
  \begin{equation}
    \label{dmft-spt}
    \Sigma^{\text{spt}}_\sigma(t,t') =
    - U(t) U(t') G_{\sigma}(t,t') G_{\bar\sigma}(t',t) G_{\bar\sigma}(t,t').
  \end{equation}
  The resulting self-consistent perturbation theory (SPT) is a truncation of
  the skeleton expansion for the self-energy, which can be derived from 
  an approximation to the Luttinger Ward-functional and is therefore
  conserving. SPT is incorporated into the DMFT iteration by replacing
  step (ii)-(iv) in Fig.~\ref{dmft-fig-calc} with a solution of the
  Dyson equation (\ref{dmft-dysonloc}) for $G_\sigma$, where
  $\Sigma_\sigma$ is given by Eq.~(\ref{dmft-ipt}). Note that in this
  implementation $G_\sigma$ is the SPT solution of the single-site
  problem for given $\Lambda$ only after the DMFT iteration is
  converged.

  When SPT is used instead of IPT as an approximate impurity solver,
  we find that $E_\text{tot}$ is indeed constant with time (solid
  lines in Fig.~\ref{dmft-fig-energy}).  However, SPT is not reliable
  at intermediate interaction strength either. For $U=5$
  (Fig.~\ref{dmft-fig-energy}b), SPT predicts a monotonous relaxation of
  $E_\text{pot}$ and $E_\text{kin}$, while the numerically exact QMC
  yields oscillations which are an important feature of the dynamics
  in the Hubbard model at strong coupling. For weak interactions, SPT
  performs slightly better, but in this parameter regime it is worse
  that the IPT solution (Fig.~\ref{dmft-fig-energy}a). The fact that
  IPT approximates the exact numerical solution better than SPT is
  already known from equilibrium DMFT.

  \section{Results}
  \label{sec-results}

  In the remainder of this paper we present additional
  numerical results for the interaction quench in the Hubbard model in
  nonequilibrium DMFT, building on our previous work
  (Ref.~\onlinecite{Eckstein2009a}).  The system is assumed to be in
  the noninteracting ground state before time $t=0$, when the
  interaction is abruptly switched to a positive value $U$. We
  consider only homogeneous nonmagnetic states at half-filling
  ($n_\uparrow=n_\downarrow=\tfrac12$). Hopping matrix elements are
  chosen such that the density of states is of semielliptic shape
  Eq.~(\ref{rho-semi}), and the quarter bandwidth $V=1$ is set as
  energy unit, so that time is measured in units of $\hbar/V=1$.

  The time evolution of various thermodynamic quantities after this
  interaction quench was already discussed in
  Ref.~\onlinecite{Eckstein2009a}. After some preliminary remarks on
  the effective temperature after a quench (Sec.~\ref{sec-teff}) we
  will briefly restate the basic conclusions of the latter publication
  and substantiate them with additional data
  (Sec.~\ref{sec-oldresults}). We then turn to a characterization of
  the relaxing state in terms of dynamical quantities, i.e., the
  spectral function (Sec.~\ref{sec-response}), and the optical
  conductivity (Sec.~\ref{sec-opt-response}).

  \begin{figure}
    \begin{center}
      \includegraphics[width=\columnwidth]{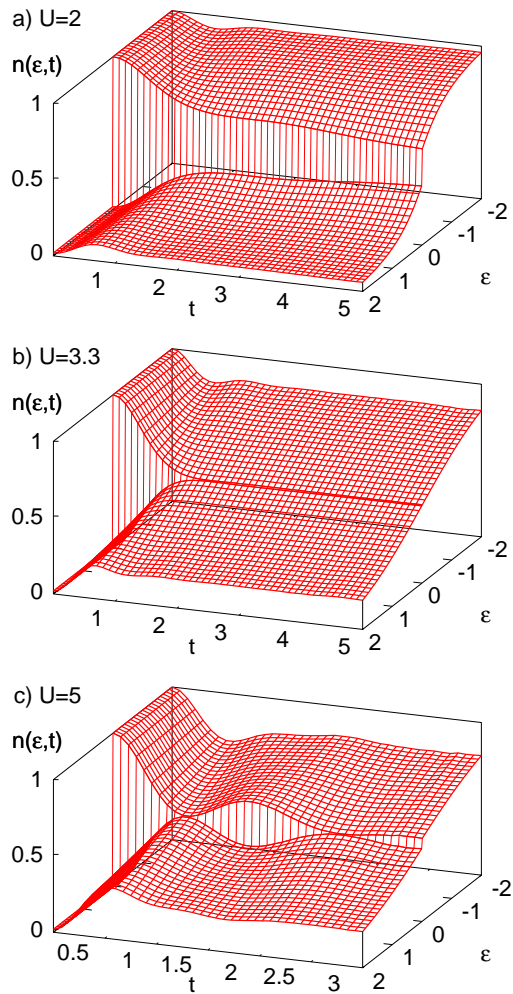}
    \end{center}
    \caption{
      Momentum distribution $n(\epsilon,t)$ after an interaction quench 
      in the Hubbard model from the noninteracting ground state to interaction 
      $U=2$ (a), $U=3.3$ (b), and $U=5$ (c).
    }
    \label{fig-nk}
  \end{figure}

  \subsection{Excitation after an interaction quench}
  \label{sec-teff}

  An important information on the state of the system after the
  interaction quench is its excitation energy with respect to the
  ground state. Because the system is assumed to be isolated from the
  environment, the total energy is conserved after the quench and its
  value follows from the expectation values of the Hamiltonian in the
  initial state immediately before the quench.  The energy corresponds
  to an effective temperature $T_\text{eff}$, i.e., the temperature of
  the unique thermal equilibrium state which has the same total energy
  [Eq.~(\ref{etot})],
  \begin{equation}
    \label{teff}
    E_\text{tot}(t)=E_\text{tot}(0^+)=
    \frac{\TR\,H e^{-H/T_\text{eff}}}
    {\TR\,e^{-H/T_\text{eff}}}.
  \end{equation}
  (An analogously defined effective chemical potential is fixed to
  $\mu_\text{eff}=0$ by particle-hole symmetry.) For the quench in the
  Hubbard model we compute $T_\text{eff}$ by a numerical solution of
  Eq.~(\ref{teff}). Thermal equilibrium expectation values of static
  quantities are obtained from equilibrium DMFT, using QMC as impurity
  solver.  For the quenches discussed below, $T_\text{eff}$ is of the
  same order as the hopping strength, which is far above the Mott
  transition endpoint in thermal equilibrium.

  If the system reaches a thermal equilibrium state a sufficiently
  long time after the quench, the temperature of this state is given
  by $T_\text{eff}$. Below we thus compare expectation values of
  observables after the quench with thermal equilibrium expectation
  values at $T=T_\text{eff}$. All static quantities in thermal
  equilibrium are directly computed within equilibrium DMFT.  The
  computation of dynamical quantities such as the spectral function
  and the optical conductivity, however, would require an analytical
  continuation from Matsubara frequencies to real frequencies, which
  is not accurate enough at large frequencies and high temperature to
  allow for a quantitative comparison. We therefore use nonequilibrium
  DMFT to obtain real-time Green functions and the real-time optical
  conductivity in thermal equilibrium directly in the time domain.
  This calculation is equivalent to an ``interaction quench'' in which
  the value of $U$ is not changed and the initial state is at finite
  temperature $T=T_\text{eff}$.  In contrast to the quench from the
  noninteracting state, it is done on the L-shaped contour, and we do
  not use the tricks which are discussed in
  Sec.~\ref{sec-flowdiagram}. The maximum times that are accessible in
  this way are comparable to the times which are accessible an the
  interaction quench from the noninteracting initial state.

  \subsection{Relaxation after an interaction quench}
  \label{sec-oldresults}

  \begin{figure}
    \begin{center}
      \includegraphics[width=0.9\columnwidth]{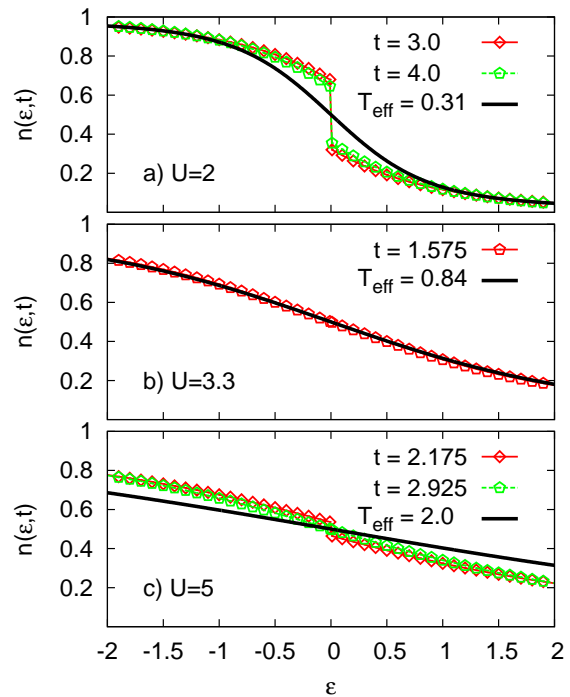}
    \end{center}
    \caption{
      Comparison of the momentum distribution $n(\epsilon,t)$ for fixed time $t$ 
      after the quench (symbols) to the momentum distribution in thermal equilibrium 
      at the effective temperature $T_\text{eff}$ [cf.~Eq.~(\ref{teff})] (solid lines).
      Interaction parameters are $U=2$ (a), $U=3.3$ (b), and $U=5$ (c).}
    \label{fig-nkthermal}
  \end{figure}
  
  \begin{figure}
    \begin{center}
      \includegraphics[width=\columnwidth]{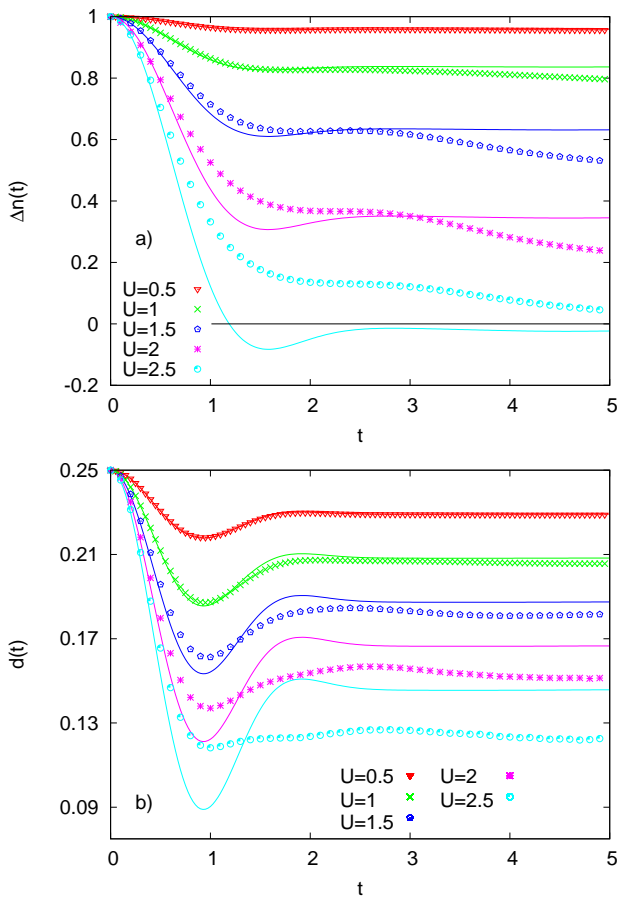}
    \end{center}
    \caption{Approach of the prethermalized state at weak-coupling and
      subsequent relaxation towards the thermal state. (a) Discontinuity
      at the Fermi surface. (b) Double occupation. Solid lines:
      weak-coupling
      results~[Eq.~(\ref{eq:npertresult})-(\ref{eq:dpertresult})].
    }\label{fig-weak-coupling}
  \end{figure}

  The time evolution after an interaction quench in the Hubbard model
  depends on the parameter $U$ in a very sensitive manner. To
  illustrate the qualitatively different relaxation behavior in the
  weak, strong, and intermediate-coupling regime we plot the momentum
  distribution $n(\epsilon,t)$ [Eq.~(\ref{neps})] for three values of
  $U$ (Fig.~\ref{fig-nk}). In all three cases the magnitude of the
  discontinuity $\Delta n(t) =
  \lim_{\eta\to0^+}[n(-\eta,t)-n(\eta,t)]$ at the Fermi energy
  decreases with time. Note that $\Delta n(t)$ remains finite for a
  finite time after the quench; for the present case of a local
  self-energy this is due to the fact that $\Delta n(t)$ is directly
  related to the retarded Green function at
  $\epsilon=0$.\cite{Eckstein2009a} Because a discontinuity in the
  momentum distribution of a Fermi liquid in thermal equilibrium can
  exist only at zero temperature, while on the other hand, a quenched
  system is always excited with respect to the ground state, the
  existence of a finite jump $\Delta n(t)$ clearly indicates that the
  system is not yet fully thermalized.  The size of the discontinuity
  is thus well suited to characterize the relaxation after the quench.

  In the weak-coupling regime (Fig.~\ref{fig-nk}a), $n(\epsilon,t)$
  rapidly evolves towards a distribution ($t\lesssim 2$ in
  Fig.~\ref{fig-nk}a) which is not yet thermalized, but changes only
  slowly in time. This emergence of long-lived nonthermal states is an
  example of prethermalization,\cite{Moeckel2008} which is observed in
  a wide range of classical and quantum systems.\cite{Berges2004a} As
  shown by Moeckel and Kehrein,\cite{Moeckel2008} the nonthermal
  state remains stable for all times within second order unitary
  perturbation theory in $U/V$, i.e., higher-order corrections become
  effective only on the long timescale $V^3/U^4$.  In the limit of
  infinite dimensions their weak-coupling result for the transient
  behavior towards the prethermalization plateau has the form
  \begin{align}
    n_{\text{pert}}(\epsilon,t)
    &=
    n(\epsilon)
    -
    4U^2F(\epsilon,t)
    \,,\label{eq:koccpretherm}\\
    F(\epsilon,t)
    &=
    \int\limits_{-\infty}^{\infty}\!dE
    \,
    \frac{\sin^2(E-\epsilon)t/2}{(E-\epsilon)^2}
    \,
    J_\epsilon(E)
    ,\label{eq:Faux}\\
    J_\epsilon(E)
    &=
    \int\!\!d\epsilon_1^\prime
    \int\!\!d\epsilon_2^\prime
    \int\!\!d\epsilon_1       
    \,
    \delta(\epsilon_1^\prime+\epsilon_2^\prime-\epsilon_1-E)
    \,
    \times\nonumber\\&
    \rho(\epsilon_1^\prime)\rho(\epsilon_2^\prime)\rho(\epsilon_1)
    \,
    [n(\epsilon)n(\epsilon_1)(1-n(\epsilon_1^\prime))(1-n(\epsilon_2^\prime))
    \nonumber\\&
    \hspace{7mm}-
    (1-n(\epsilon))(1-n(\epsilon_1))n(\epsilon_1^\prime)n(\epsilon_2^\prime)]
    .
  \end{align}
  For a half-filled band and a symmetric density of states,
  $\rho(\epsilon)$ $=$ $\rho(-\epsilon)$, we obtain
  \begin{align}
    F(\epsilon,t)
    &=
    -
    \frac{\text{sgn}(\epsilon)}{2}
    \,
    \int\limits_{0}^{t}\!
    ds
    \,
    (t-s)
    \,
    \text{Re}\big[
    R(s)^3
    \,
    e^{is|\epsilon|}
    \big]
    ,
  \end{align}
  where $R(s)$ $=$
  $\int\!d\epsilon\,\Theta(-\epsilon)\,\rho(\epsilon)\,e^{is\epsilon}$.
  This yields $\Delta n(t)$ and also $d(t)$ by using the energy
  conservation after the quench,
  \begin{align}
    \Delta n_{\text{pert}}(t)
    &=
    1-4U^2
    \int\limits_{0}^{t}\!
    ds
    \,
    (t-s)
    \,
    \text{Re}\big[R(s)^3\big]
    ,\label{eq:npertresult}\\
    d_{\text{pert}}(t)
    &=
    \frac{1}{4}
    -
    2U
    \int\limits_{0}^{t}\!\!
    ds
    \,
    \text{Im}\big[R(s)^4\big]
    .\label{eq:dpertresult}
  \end{align}
  Numerical evaluations of these functions are plotted and compared to
  our DMFT results in Fig.~\ref{fig-weak-coupling} for the
  semielliptic density of states (\ref{rho-semi}) with $V=1$.
  Regarding the transient behavior and the prethermalization plateau
  we find very good agreement for $U\lesssim 1$. Interestingly the
  prethermalization plateau of $\Delta n(t)$ is almost correctly predicted by the
  weak-coupling results even for $U\lesssim 2$.  For larger times the
  system relaxes further towards the thermal value.

  In the strong-coupling regime (Fig.~\ref{fig-nk}c), the relaxation
  is dominated by damped collapse and revival oscillations of
  approximate periodicity $2\pi/U$. The decay of these oscillations is
  not fully accessible within CTQMC due to the dynamical sign problem.
  However, our results show that $n(\epsilon,t)$ oscillates around a
  nonthermal distribution (Fig.~\ref{fig-nkthermal}c). This behavior,
  which is analogous to prethermalization at weak-coupling, is similar
  to what was found for the double occupation
  $d(t)$,\cite{Eckstein2009a} i.e., a decay on the timescale $1/V$ to
  oscillations around a nonthermal value which does not change on much
  longer timescales.

  The interaction quench to $U=3.3V$ is characterized by a rapid
  thermalization of the momentum distribution (Figs.~\ref{fig-nk}b and
  \ref{fig-nkthermal}b), without signatures of either collapse and
  revival oscillations or a prethermalization plateau in $n(\epsilon,t)$. 
  Numerically we cannot detect a finite width to
  the crossover regime between the weak- and strong-coupling behavior,
  which indicates that there is a single point $U=\Ucdyn \approx
  3.2V$ which marks a {\em dynamical transition} in the Hubbard
  model.\cite{Eckstein2009a} A further investigation of this
  phenomenon and its relation to the Mott transition in equilibrium
  will require a systematic analysis of interaction quenches which
  start from a wide range of initial states other than the
  noninteracting ground state. This is left to a future publication.
  In the following we turn to a different question and investigate to
  what extent the rapid thermalization close to $U = \Ucdyn$, the
  oscillations at $U>\Ucdyn$, and the prethermalization at $U<\Ucdyn$
  show up in various dynamical quantities of the Hubbard model.

  \begin{figure}
    \begin{center}
      \includegraphics[width=\columnwidth]{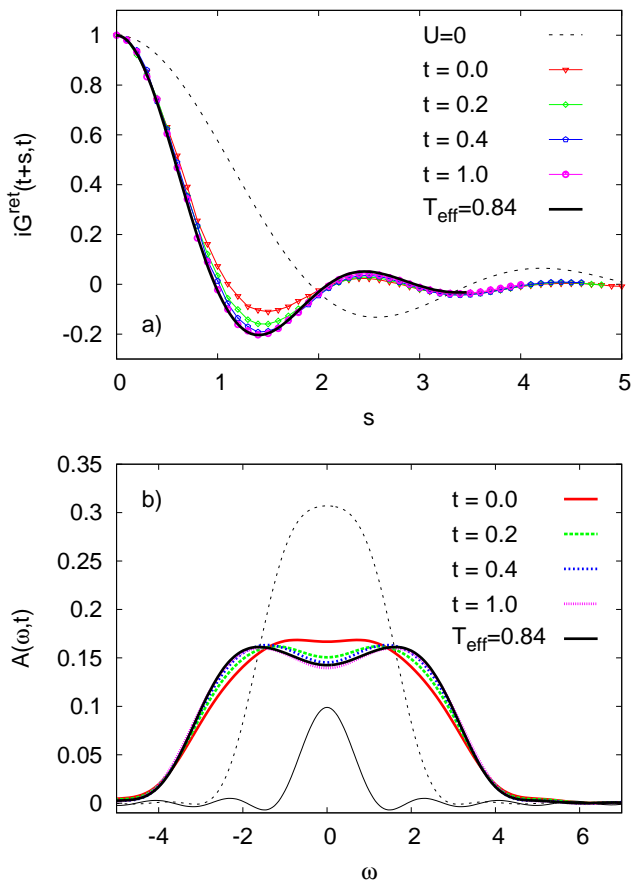}
    \end{center}
    \caption{
      (a) Local Green function $G^\ret(t+s,t)$ for an
      interaction quench in the Hubbard model to $U=3.3$ 
      (slightly above $U_\text{dyn}$). The function
      is purely imaginary due to particle-hole symmetry. The solid
      black line is the Green function in the thermal equilibrium
      state ($U=3.3$, $T_\text{eff}=1/\beta=0.84$).  The dotted line
      ($U=0$) is the retarded Green function in the noninteracting
      initial state.
      (b) Spectral function $A(\omega,t)=-(1/\pi) \text{Im}
      G^\ret(\omega,t)$ for the same parameters as in the upper panel.
      The dotted line and the line labelled $T_\text{eff}=0.84$ are
      the semielliptic density of states~(\ref{rho-semi})
      of the initial state and the
      thermal equilibrium spectrum at temperature $T_\text{eff}=0.84$,
      respectively.  Spectra are obtained from Fourier transformation
      of real-time quantities, and the Fourier integral
      (\ref{partial-ft}) is cut off at $s_\text{max}=3.5$ with an
      additional Gaussian factor (see text). The
      corresponding kernel [Eq.~(\ref{spectrum-kernel}), $\kappa=0.1$]
      is shown as thin solid line.
    }
    \label{fig-spectrum1}
  \end{figure}

  \subsection{Spectral function}
  \label{sec-response}

  Important information about a correlated system out of equilibrium
  cannot only be obtained from thermodynamic quantities, but also from
  the dynamical response of the system to certain external
  perturbations, which can be computed from various real-time
  correlation functions. In the following subsection we discuss the
  time evolution of the local Green function $G_\sigma(t,t')\equiv
  G(t,t')$ in the paramagnetic phase of the Hubbard model after a
  quench from the noninteracting ground state to finite interaction
  $U$.  For this purpose we introduce the partial Fourier transform
  \begin{equation}
    \label{partial-ft}
    G^{\ret,\les}(\omega,t) = \int \! ds\,e^{i \omega s} G^{\ret,\les}(t+s,t)
  \end{equation}
  of the retarded and lesser Green function, and the spectral function
  $A(\omega,t)=-(1/\pi)\,\text{Im}\,G^\ret(\omega+i0,t)$. The spectrum
  turns out to be a useful representation of the nonequilibrium Green
  function, although it lacks a direct relation to the ``distribution
  function'' $G^{\les}(\omega,t)$ and thus does not have the same
  significance as in the equilibrium case. [In equilibrium one has
  $G^{\les}(\omega)=2\pi i A(\omega)f(\omega)$.]

  Before discussing the results we have to mention a technicality,
  which arises from the restriction of the Monte Carlo simulations to
  relatively small times $t<\tmax$. In practice, the integration range
  in Eq.~(\ref{partial-ft}) must be cut off at $s_\text{max}\equiv
  \tmax-t$, leading to artificial oscillations at frequency
  $1/s_\text{max}$. To reduce this effect in a controlled way we
  introduce an additional Gaussian factor $\exp(-s^2\kappa)$ in the
  integral (\ref{partial-ft}).  The resulting expression amounts to a
  convolution of the true Fourier transform ($\tmax=\infty$) with the
  kernel
  \begin{equation}
    \label{spectrum-kernel}
    k(\omega;\kappa,s_\text{max})=
    \frac{1}{2\pi}
    \int\limits_{-s_\text{max}}^{s_\text{max}} \!\!\!ds\,\exp(i\omega s-s^2\kappa).
  \end{equation}
  A suitable choice of the parameter $\kappa$ can in some cases
  suppress the oscillations without washing out important spectral
  features, and a comparison with a known equilibrium spectrum is
  always possible without loss of information after convolution of the
  latter with the same kernel.

  In Fig.~\ref{fig-spectrum1} we plot $G^\ret(t+s,t)$ and
  $A(\omega,t)$ for a quench to interaction $U=3$. The spectrum
  $A(\omega,t)$ differs from the initial semielliptic density of
  states for all times $t \ge 0$, because the choice of the Fourier
  transform in Eq.~(\ref{partial-ft}) implies that the initial
  equilibrium Green function does not enter the definition of
  $A(\omega,t)$ for $t>0$. Note that this would be different for the
  common definition of the Fourier transform at constant average time
  $(t+t')/2$.\cite{Turkowski2005a} Within numerical accuracy, both
  $G^\ret(t+s,t)$ and $A(\omega,t)$ become time-($t$)-independent for
  $t > 1/V$. This timescale is comparable to the relaxation time of
  the double occupation and the momentum distribution at $U\approx3.3$
  (Fig.~\ref{fig-nk}b).

  An important interpretation of the finite relaxation time in
  $A(\omega,t)$ can be inferred directly from the definition of the
  Green function.  According to Eq.~(\ref{ky-cntr-ret1}),
  $G^\ret(t+s,t)$ is related to the survival amplitude of local
  single-particle excitations which are created at time $t$ and
  destroyed at later time $t+s$. The decay of such an excitation
  depends on both the Hamiltonian, which defines the possible
  scattering mechanisms, and the quantum state of those particles
  which act as scatterers.  While the Hamiltonian changes abruptly at
  $t=0$, the latter evolves with time, leading to the finite
  relaxation time of $A(\omega,t)$.  In contrast, $A(\omega,t)$ would
  be constant immediately after a quench in a noninteracting system,
  because the anticommutator in Eq.~(\ref{ky-cntr-ret1}) is a c-number
  for a quadratic Hamiltonian.  We can thus conclude that the finite
  relaxation time observed in Fig.~\ref{fig-spectrum1} is a true
  many-body effect, in analogy to the well-known fact that equilibrium
  spectra depend on temperature only for interacting systems.

  To characterize the final state after the relaxation, its spectrum
  should be compared to the equilibrium spectrum of a correlated metal
  at rather high temperature. In fact, $A(\omega,t)$ is strongly
  modified with respect to the semielliptic density of states, with
  precursors of the Hubbard bands around $\omega=\pm 2$.  The fact
  that the spectrum is not pinned at $\omega=0$ can be attributed to
  the strong excitation of the system with respect to the ground
  state. A quantitative analysis of the spectrum requires the
  knowledge of the equilibrium spectrum at the effective temperature
  $T_\text{eff}$ [cf.~Eq.~(\ref{teff}), $T_\text{eff}=0.84$ for
  $U=3.3$].  Equilibrium spectra are usually computed from
  imaginary-time correlation functions using (maximum entropy)
  analytical continuation, which is not accurate enough at high
  frequencies to allow for a comparison of two rather similar spectra.
  Using nonequilibrium DMFT, however, we can avoid this complication
  and compute real-time {\em equilibrium} Green function
  $G^\ret_{eq}(t,t') \equiv g^\ret(t-t')$ without analytical
  continuation (cf.~Sec.~\ref{sec-teff}). Within numerical accuracy,
  the resulting equilibrium function indeed agrees with the retarded
  Green function $G^\ret(t,t')$ after relaxation
  (Fig.~\ref{fig-spectrum1}a), which proves that the rapid
  thermalization at $U \approx 3.3$ can also be seen in the spectral
  function.

  \begin{figure}
    \begin{center}
      \includegraphics[width=1\columnwidth]{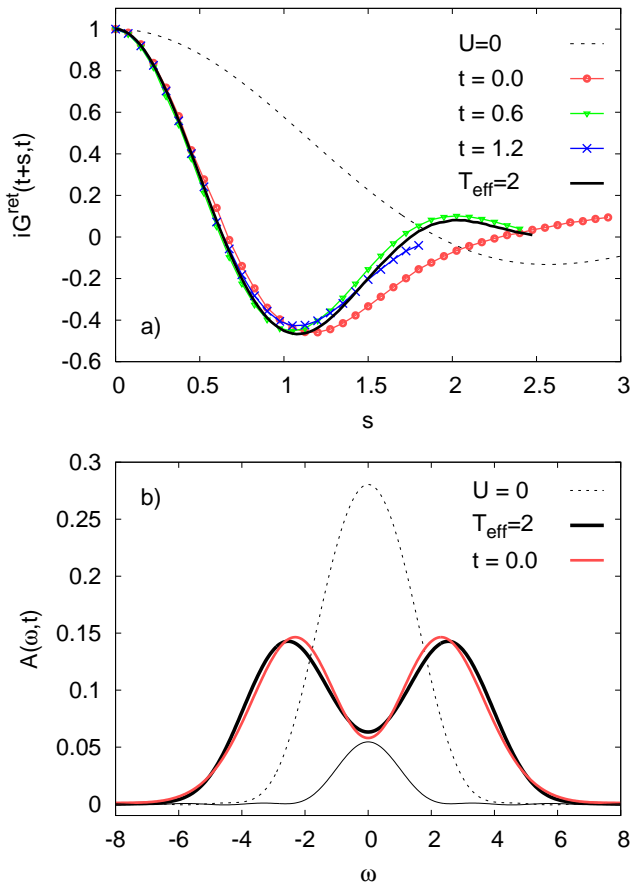}
    \end{center}
    \caption{
      Same as Fig.~\ref{fig-spectrum1}, but for the interaction quench
      to $U=5$.  Spectra (b) are obtained from Fourier transformation
      of real-time quantities, and the Fourier integral
      (\ref{partial-ft}) is cut off at $s_\text{max}=2.5$ with an
      additional Gaussian factor (see text). The corresponding kernel
      [Eq.~(\ref{spectrum-kernel}), $\kappa=0.4$] is shown as thin
      solid line.
    }
    \label{fig-spectrum2}
  \end{figure}

  The analysis of the spectrum can now be repeated for quenches to the
  weak- and strong-coupling regime. For $U \ll V$, however, the
  spectrum remains close to the semielliptic density of states for all
  times, such that rather high numerical accuracy would be needed for
  a systematic investigation of the small differences. In the
  strong-coupling regime, on the other hand, the restriction to small
  times $t<\tmax$ turns out to be more limiting for an investigation
  of the retarded Green function than for static quantities, simply
  because $G^\ret(t+s,t)$ is known only for $t<\tmax-s$ and not for
  $t<\tmax$. Nevertheless, one can see that the relaxation of the
  Green function after a quench to $U=5$ (Fig.~\ref{fig-spectrum2}a)
  roughly follows the oscillatory behavior of the momentum
  distribution (Fig.~\ref{fig-nk}c): Close coincidence with the
  thermal function is reached around the time when the jump of the
  momentum occupation has its first minimum ($t=0.6$), after which the
  deviations to the thermal Green function slightly increase again.
  Similar behavior was found for the double occupation, which comes
  closest to the thermal value at its first minimum around
  $t=0.6$.\cite{Eckstein2009a} In spite of the large effective
  temperature ($T_\text{eff}=2V$), the spectral function has a clear
  minimum at $\omega=0$, and well-pronounced Hubbard bands at
  $\omega\approx\pm U/2$ (Fig.~\ref{fig-spectrum2}b). However, the
  absolute changes with time are small in the strong-coupling regime.
  This behavior is expected because it can be shown that the spectrum
  is independent of time $t$ after a quench to the atomic limit.  

  \subsection{Optical conductivity}
  \label{sec-opt-response}

  \begin{figure}
    \begin{center}
      \vspace*{-5mm}
      \hspace*{-5mm}\includegraphics[width=1.2\columnwidth]{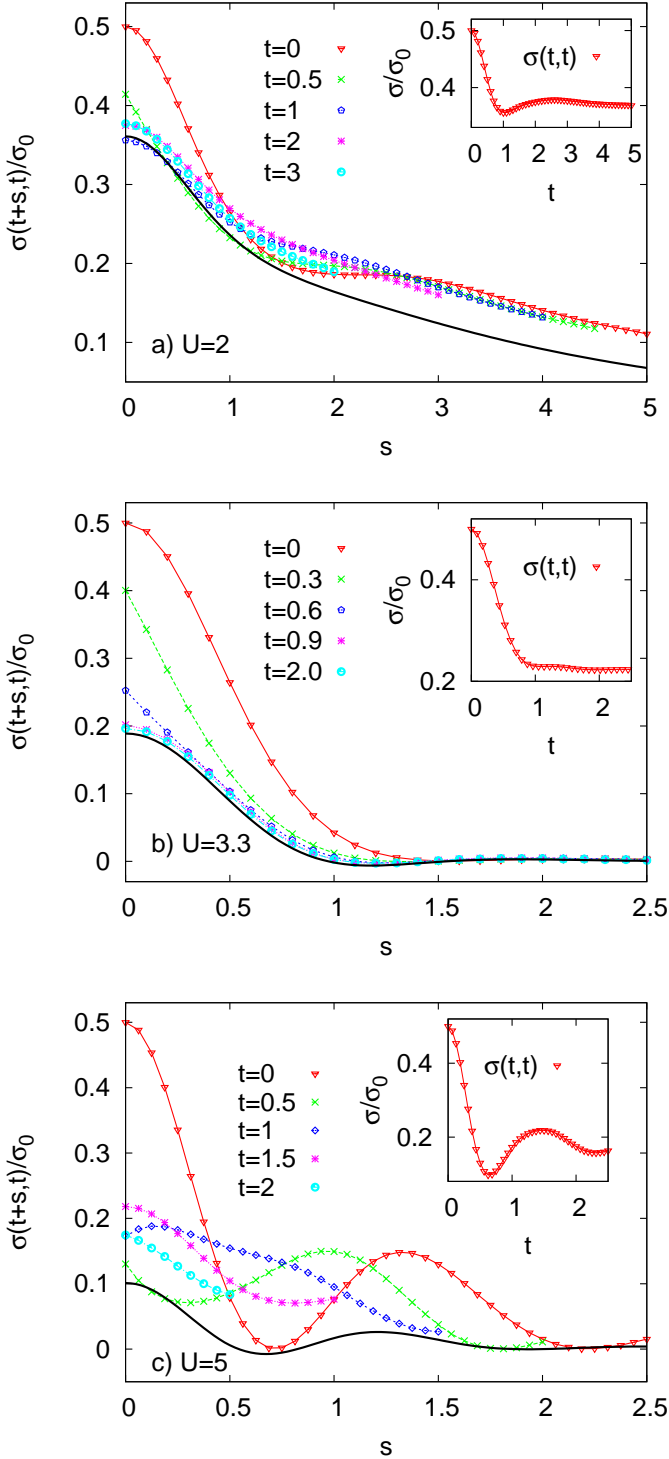}
    \end{center}
    \caption{
      Optical conductivity $\sigma(t+s,t)$ [Eq.~(\ref{conductivity})] after 
      quenches to $U=2$ (a), $U=3.3$ (b), and $U=5$ (c). The inset shows
      $\sigma(t,t)$, and black solid lines correspond to the optical conductivity
      in thermal equilibrium at $T_\text{eff}=0.37$ (a), $T_\text{eff}=0.84$ (b), and
      $T_\text{eff}=2$ (c).
    }
    \label{fig-opt}
  \end{figure}

  The two-time optical conductivity $\sigma(t,t')$ describes the
  linear response of the electrical current in a nonequilibrium state
  to a time-dependent electrical field $\delta{\bm E}(t)$ (which we
  call the probe field),
  \begin{equation}
    \label{conductivity}
    \delta \expval{{\bm j}(t)} = \int\limits_{-\infty}^t \!\!\! d\tb\, \sigma(t,\tb) \delta{\bm E}(\tb).
  \end{equation}
  (Tensor notation of $\sigma(t,t')$ is suppressed.) In solids,
  optical spectroscopy on nonequilibrium states is usually performed
  within the pump-probe setup, where the system is driven out of
  equilibrium by a strong laser pulse (the pump). In the following we
  calculate $\sigma(t,t')$ after the interaction quench to see how the
  electrical response becomes stationary while the system relaxes
  towards its thermal equilibrium state.

  Microscopically, the optical conductivity is related to the
  current-current correlation function, which can be computed from two
  diagrammatic contributions: (i) The bubble diagram of two Green
  functions $G_\Vk$ and the current vertex $v_\Vk=\partial
  \epsilon_\Vk/\partial \Vk$, and (ii) diagrams containing the vertex
  corrections of the current vertex.\cite{Pruschke1993a} Within
  equilibrium DMFT, vertex corrections are local and thus do not
  contribute to the conductivity because $v_\Vk$ is antisymmetric
  under inversion of $\Vk$, and $G_\Vk$ is
  symmetric.\cite{Khurana1990a} In a nonequilibrium situation these
   conditions can be violated, e.g., due to an electrical pump field,
   in which case the conductivity depends on the relative polarization
   of pump and probe, so that vertex corrections do
   contribute.\cite{Eckstein2008b,Tsuji2009a} However, for the
   interaction quench the inversion symmetry of the state is preserved,
   and $\sigma(t,t')$ can be calculated from the bubble diagram
   alone.\cite{Eckstein2008b}

  The microscopic derivation of $\sigma(t,t')$ within nonequilibrium
  DMFT was discussed in detail in Ref.~\onlinecite{Eckstein2008b}. 
  In  the following we thus only state the results for $\sigma(t,t')$
  after an interaction quench in the Hubbard model on the hypercubic
  lattice in $d=\infty$, with hopping amplitudes that yield a
  semielliptic density of states\cite{Bluemer2003a}
  (Eq.~(\ref{rho-semi}) with $V=1$, as above). The band dispersion
  $\epsilon_\Vk$ enters the expression via the current vertex
  $v_\Vk=\partial \epsilon(\Vk)/\partial \Vk$; this is where the
  hopping amplitudes enter in addition to the density of states.
  Conductivity is measured in units of $\sigma_0$ $=$ $2\rho a^2 e^2 V
  /\hbar^2$, where $a$ is the lattice constant, and $\rho$ is the
  number of lattice sites per volume.

\begin{figure}
    \begin{center}
      \vspace*{0mm}
      \includegraphics[width=1\columnwidth]{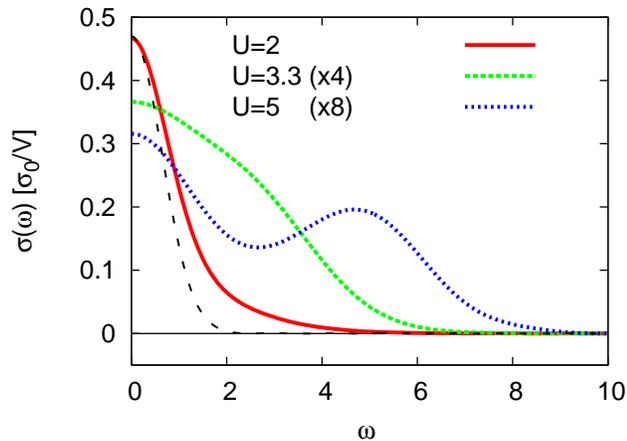}
    \end{center}
    \caption{
      Frequency-dependent optical conductivity in thermal equilibrium
      at the temperature $T_\text{eff}=0.37$ ($U=2$),
      $T_\text{eff}=0.84$ ($U=3.3$), and $T_\text{eff}=2$ ($U=5$). The
      latter two curves are scaled by a factor $4$ and $8$,
      respectively. All curves were obtained by Fourier transformation
      (\ref{sigma-ft}) of real-time data. In the case of $U=2$, the
      Fourier integral is cut off at $s_\text{max}=5$ with an
      additional Gaussian factor, as explained for the
      spectral function. The corresponding kernel
      [Eq.~(\ref{spectrum-kernel}), $\kappa=0.2$]
      is shown as dotted line.
    }
    \label{fig-opt-eq}
  \end{figure}

  In Fig.~\ref{fig-opt}, $\sigma(t+s,t)$ is plotted as a function of
  time-difference $s$. This parametrization is most convenient for
  analyzing how the electrical response of the system becomes
  stationary (i.e., independent of $t$) during the relaxation.  The
  results are compared to the optical conductivity
  $\sigma_\text{eq}(s)$ in thermal equilibrium, which is obtained
  directly from nonequilibrium DMFT without analytical continuation
  (cf.~Sec.~\ref{sec-teff}). The more familiar frequency-dependent
  optical conductivity
  \begin{equation}
    \label{sigma-ft}
    \sigma_\text{eq}(\omega) = \text{Re} \int\limits_0^\infty \!ds\,
    e^{i\omega s} \sigma_\text{eq}(s) 
  \end{equation}
  is plotted in Fig.~\ref{fig-opt-eq}.

  After quenches to weak-coupling ($U=2$, Fig.~\ref{fig-opt}a),
  $\sigma(t,t')$ undergoes a rapid initial relaxation, but it does not
  approach the thermal value within the accessible times. This
  behavior reflects the prethermalization that is observed in the
  momentum occupation. The conductivity at the corresponding effective
  temperature ($T_\text{eff}=0.37$) consists of a Drude peak at
  $\omega=0$ (Fig.~\ref{fig-opt-eq}), which is only slightly broadened
  due to temperature and interaction. Because a narrow Drude peak
  implies a slow decay of $\sigma_\text{eq}(s)$ with time difference,
  we cannot resolve the true width of the peak from data which are
  restricted to small times.

  For a quench to $U=3.3$ (Fig.~\ref{fig-opt}b), we observe a rapid
  relaxation of the optical response. The optical conductivity depends
  only on time difference for $t\gtrsim 1/V$ and coincides with
  $\sigma_\text{eq}(s)$ for the effective temperature
  $T_\text{eff}=0.67$. The latter falls off rather quickly with time
  differences $s$, indicating that the Drude peak is strongly broadened
  because of the large temperature and the relatively strong
  interaction (Fig.~\ref{fig-opt-eq}).

  Finally, for the quench to $U=5$ (Fig.~\ref{fig-opt}c) relaxation to
  the thermal state becomes again slower than at $U=3.3$. We observe
  the characteristic collapse and revival oscillations when
  $\sigma(t+s,t)$ is plotted at fixed time difference $s$ (inset in
  Fig.~\ref{fig-opt}c). Due to the large effective temperature
  ($T_\text{eff}=2$) the conductivity of the corresponding equilibrium
  state is rather a bad metal than an insulator, but nevertheless the
  Hubbard band at $\omega=U$ is clearly separated from the broad
  feature at $\omega=0$ (Fig.~\ref{fig-opt-eq}).

  \section{Conclusion}

  In this paper we described in detail how weak-coupling
  continuous-time quantum Monte Carlo (QMC) can be used as an impurity
  solver within nonequilibrium DMFT. The formalism, which was used in
  Ref.~\onlinecite{Eckstein2009a} to investigate the interaction
  quench in the Hubbard model, was extended to the case when the
  initial state is a finite temperature equilibrium state at nonzero
  interaction $U$. Because nonequilibrium experiments in interacting
  systems often start from correlated initial states rather than the
  noninteracting ground state, this extension is a prerequisite to
  apply DMFT within a variety of experimental situations in the field
  of cold atomic gases and time-resolved spectroscopy on correlated
  solids.

  We used the numerically exact QMC solution of the DMFT equations to
  benchmark the generalization of the iterated perturbation theory
  (IPT) to the Keldysh contour.  We find that IPT is remarkably good
  at weak interactions. However, in contrast to the equilibrium case
  it yields unphysical results in the intermediate-coupling regime and
  thus cannot provide a reasonable interpolation between the weak- and
  strong-coupling regime. The reason is that IPT is not a conserving
  approximation, which can lead to an explicit violation of the energy
  conservation as a function of time in some parameter regime.

  Furthermore, we used the nonequilibrium formalism to solve a system
  in thermal equilibrium. In this way one can avoid analytical
  continuation and obtain dynamical quantities in real time instead of
  imaginary time. We used this approach to compute the spectral
  function and the optical conductivity of the single-band Hubbard
  model. Due to the dynamical sign problem of QMC one is restricted to
  relatively short times, such that frequency-dependent quantities,
  which are obtained from real-time functions by Fourier
  transformation, are considerably broadened. The real-time formalism
  can thus not directly replace the conventional analytical continuation from
  Matsubara to real frequencies. However, since the kernel which
  mediates the broadening of the spectra is explicitly known, it 
  may be useful either to judge the accuracy of
  analytically continued spectra, or improve the analytical
  continuation in some frequency range.

  In the last part of this paper we presented further results for the
  interaction quench in the Hubbard model. In particular, we
  investigated the time evolution of the real-time Green functions. It
  was shown that the different relaxation behavior at weak, strong and
  intermediate coupling, which was characterized by the time evolution
  of the double occupation and the momentum distribution in
  Ref.~\onlinecite{Eckstein2009a}, is also reflected in the
  nonequilibrium spectral function: In the weak- and strong-coupling
  regime a thermal state cannot be reached within the accessible
  times, whereas the spectrum (as well all quantities that can be
  obtained from it) rapidly relaxes to the thermal equilibrium at
  intermediate coupling ($U=\Ucdyn$).

  The fact that the very sensitive $U$-dependence of the relaxation
  behavior is manifest also in the spectral function suggests that the
  phenomenon of fast electronic thermalization near $\Ucdyn$ may also
  be observed with pump-probe spectroscopy on correlated systems.
  Further details of this transition-like phenomenon will hopefully
  soon be clarified by means of the DMFT+QMC formalism presented in
  this work.


  \section*{Acknowledgements}

  M.E.\ acknowledges support by Studienstiftung des deutschen Volkes.
  This work was supported in part by the SFB 484 of the Deutsche
  Forschungsgemeinschaft (DFG) and the Swiss National Science
  Foundation (PP002-118866). CTQMC calculations were run on the Brutus
  cluster at ETH Zurich, using the ALPS library.\cite{ALPS}


\end{document}